\documentclass[aps,twocolumn,superscriptaddress,nofootinbib]{revtex4}

\usepackage{amsfonts}
\usepackage{amssymb}
\usepackage{mathdots}

\usepackage{graphicx}
\usepackage{bbm}
\usepackage{color}
\usepackage{pifont}
\usepackage{enumerate}
\usepackage{subfigure}
\usepackage{color}
\newcommand*{\rom}[1]{\expandafter\@slowromancap\romannumeral #1@}
\makeatother
\usepackage{mathptmx}
\usepackage{mathrsfs}
\usepackage{mathtools}
\usepackage{amsmath}
\usepackage{amsfonts}
\usepackage{amssymb}
\usepackage{mathdots}
\usepackage{bbold}
\usepackage{isomath}
\usepackage{mathrsfs}
\usepackage{amsthm} 
\usepackage{graphicx}
\usepackage{bbm}
\usepackage{color}
\usepackage{pifont}
\usepackage{enumerate}
\usepackage{amsthm}
\theoremstyle{plain}
\theoremstyle{definition}
\usepackage{mathtools}
\makeatletter
\newsavebox{\@brx}
\newcommand{\llangle}[1][]{\savebox{\@brx}{\(\m@th{#1\langle}\)}%
  \mathopen{\copy\@brx\kern-0.5\wd\@brx\usebox{\@brx}}}
\newcommand{\rrangle}[1][]{\savebox{\@brx}{\(\m@th{#1\rangle}\)}%
  \mathclose{\copy\@brx\kern-0.5\wd\@brx\usebox{\@brx}}}
\makeatother
\usepackage[titletoc]{appendix}
\usepackage[english]{babel}
\usepackage[autostyle]{csquotes}
\usepackage{hyperref} 
\usepackage{cleveref} 

\DeclareMathOperator{\sech}{sech} 

\begin{document}

\title{Adiabatic-NOVEL for Nano-Scale Magnetic Resonance Imaging}



\author{R. Annabestani}
 \email[]{rannabes@uwaterloo.ca}
\affiliation{Institute for Quantum Computing, University of Waterloo, Waterloo, Ontario N2L 3G1, Canada}
\affiliation{Department of Physics and Astronomy, University of Waterloo, Waterloo, Ontario N2L 3G1, Canada}
\author{M. S. Mirkamali}
\email[]{msmirkam@uwaterloo.ca}
\affiliation{Institute for Quantum Computing, University of Waterloo, Waterloo, Ontario N2L 3G1, Canada}
\affiliation{Department of Physics and Astronomy, University of Waterloo, Waterloo, Ontario N2L 3G1, Canada}
\author{R. Budakian}
\email[]{rbudakian@uwaterloo.ca}
\affiliation{Institute for Quantum Computing, University of Waterloo, Waterloo, Ontario N2L 3G1, Canada}
\affiliation{Department of Physics and Astronomy, University of Waterloo, Waterloo, Ontario N2L 3G1, Canada}



\begin{abstract}
 We propose a {\color{black}highly} efficient dynamic nuclear polarization technique that is robust against field in-homogeneity. This technique is designed to enhance the detection sensitivity in nano-MRI, where large Rabi field gradients are required. The proposed technique consists of an adiabatic half passage pulse followed by an adiabatic linear sweep of the electron Rabi frequency and can be considered as an adiabatic version of nuclear orientation via electron spin locking (adiabatic-NOVEL). We analyze the spin dynamics of an electron-nuclear system that is under microwave irradiation at high static magnetic field and at cryogenic temperature. The result shows that an amplitude modulation of the microwave field makes adiabatic-NOVEL highly efficient and robust against both the static and microwave field in-homogeneity.
\end{abstract}

\maketitle


\section{introduction}
 Dynamic nuclear polarization (DNP) is a powerful technique in nuclear magnetic resonance (NMR) spectroscopy that enhances the signal to noise ratio. DNP transfers the magnetization from highly polarized spins, such as electrons, to spins with low polarization, such as protons, or from abundant spin species to dilute spin species \cite{A62, S96,S65}. Since the early invention of DNP in 1950s \cite{Over53, CS53}, there have been extensive applications such as reactions in bio-molecular chemistry and enhancing image contrast in MRI  \cite{Gol06,Gol062, Sche17, Lon13}.  
 
 Magnetic resonance imaging (MRI) is a spin based spectroscopy technique that is widely used for studying molecular, chemical and structural properties of biological systems such as human body \cite{A09, Nelson08}. There have been a significant interest to extend the MRI capability to nanometer scale (nano-MRI) in order to obtain information about nano-size biological structures such as viruses and inter-cell elements \cite{M08, T08, MM10, Ru04,De09,N13, B99, KL04, Sa11}. Despite considerable advancements, it is still a significant challenge in the field to obtain a high resolution image of nanometer size spin ensemble, due to both relatively low magnetic moment of nuclear spins and small sample size. One way of increasing the detection sensitivity of nano-MRI is to incorporate a novel DNP method to initialize a nanometer size spin ensemble and use it as a sample preparation step for the imaging \cite{Nelson08}. Since large field gradients are often an integral part of the imaging process, the DNP technique of interest needs to be robust against magnetic field variation across the sample.  
 
 %
  Microwave induced DNP methods, regardless of their diversity, can be categorized into two groups: Incoherent processes, such as Overhauser effect and solid effect \cite{Over53,CS53, HS68}, and coherent processes, such as the cross polarization (CP) between unlike nuclear spins \cite{P73,M79} and the electron-nuclear cross polarization (eNCP) \cite{W06}. In an incoherent DNP process, a continuous microwave irradiation with the assistance of thermal relaxations drives the electron-nuclear spin system towards a new quasi-equilibrium state where the nuclear spin signal is ideally enhanced by a factor of $\gamma_{e}/ \gamma_{n}$ (= 660 in case of protons). In practice, the microwave irradiation may simultaneously excite various DNP pathways that polarize the nuclear spin in opposite directions. As a result, the net DNP enhancement is suppressed. One solution is to increase the static magnetic field to isolate the competing DNP pathways. However, that give rise to a second challenge \cite{Sara13}. The microwave induced transition rates are in inverse relation with the static field strength, such as $B_{0}^{-2}$ in case of solid effect and as $B_{0}^{-1}$ in case of thermal mixing effect \cite{W06, YFS10}. Therefore, increasing the static field reduces the efficiency of incoherent DNP processes. Moreover, the rate of polarization buildup can be very slow, because the microwave induced saturation rates are competing with the thermal relaxation rates.

In a coherent DNP process the frequency and the amplitude of the microwave and/or RF fields are controlled such that the energy of the polarizing agent in the rotating frame of the pulse becomes on resonance with that of the target spin. This leads to an optimal coherent polarization exchange process between the two spins, that is mediated by their spin-spin interaction. In 1962, Harman and Hahn pioneered this \textit{energy matching} DNP mechanism \cite{H62}, and since then, the method has been adopted in several experiments, such as nuclear orientation via electron spin locking (NOVEL) \cite{HDSW88,B87, H08, W06, P73,M79}. The polarization transfer of a coherent DNP process can occur very quickly. In NOVEL, for instance, the transfer rate is determined by the inverse of the hyperfine coupling strength, which can be in the order of tens of MHz. However, NOVEL is very sensitive to deviations from the energy matching condition. {\color{black}In particular, when there is a significant field distribution across the spin ensemble, as  often is the case in nano-MRI, only a fraction of electrons would effectively exchange their polarization with their surrounding nuclear spins. As a result, the net DNP enhancement is significantly reduced.}
  
 
In this work, we look for a fast coherent DNP method to hyperpolarize a nano-scale spin ensemble to use as an initialization step for magnetic resonance force detection microscopy (MRFM). In this nano-MRI technique, the interaction between a time-dependent magnetic field gradient and the spin ensemble creates an oscillating force, that is detected through an opto-mechanical system \cite{SGBRZ95}. This device can operate at high field (a few Tesla) and at cryogenic temperature (300 mK), where the electron's state is almost fully polarized. In addition, it generates very strong time dependent microwave field (1-100 MHz) at high frequency (MHz-GHz). Therefore, the MRFM device of interest is an ideal test bed for implementing NOVEL at high field to achieve hyperpolarization. However, since large field gradients ($10^6$ T/m) are the integral components of the imaging process, the conventional NOVEL is not a suitable candidate{\color{black}, but a modification of that can be.} 


We propose an adiabatic version of NOVEL (\textit{adiabatic-NOVEL}), where the amplitude of the Rabi frequency is linearly modulated over a wide range. We analyze the spin dynamics of an electron-nuclear coupled system for two cases of a constant and a time-varying Rabi field. We conclude that an amplitude modulation of the microwave field makes the proposed DNP technique highly efficient and very robust against field in-homogeneity. We show that the adiabatic-NOVEL, which is designed primarily for dealing with field gradients, dramatically improves the performance of NOVEL even in the absence of any field distribution. 

The idea of modulating the amplitude of the irradiating field has been studied both theoretically and experimentally for the case of unlike nuclear spins \cite{P57,Pro62,Can15}. However, it has not been explored for the electron-nuclear case, until very recently \cite{Can15, Can17}. The authors well explained the high efficiency of adiabatic-NOVEL, and as complementary, our work analyzes the robustness of this DNP technique against field in-homogeneity which is of great importance for nano-MRI applications.


\section{Experiment}
\label{Sec_Experiment_Layout}

In this section, we provide an overview of the hyperpolarization experiment, and later in section. \ref{Sec_theory}, we analyze the spin dynamics and elaborate on the underlying physics. 

Consider an electron-nuclear spin system in the presence of a static magnetic field, $B_{0}\ \hat{z}$. In nuclear orientation via electron spin locking (NOVEL) method, a transverse microwave field, $B_{1} \ \hat{x}$ is irradiated on resonance with the electron Zeeman frequency, \textit{locking} the electron to the rotating frame of the pulse. The intensity of the pulse is chosen such that the electron Rabi frequency ($\omega_{1e} = \gamma_{e} B_{1}$)  in the rotating frame is on resonance with the nuclear larmor frequency  ($\omega_{0n} = \gamma_{n} B_{0}$) in the lab frame. Thus,   
 
\begin{equation}
\label{Eq_HH}
\gamma_{e} B_{1}= \gamma_{n} B_{0}
\end{equation}
Under the above \textit{energy matching condition}, the electron comes in \textit{contact} with the coupled nuclear spin. As a result the polarization exchange rate between the the two spins, that is due to their spin-spin interaction, is optimized. This electron-nuclear energy matching condition in a single spin rotating frame resembles the well known Hartman-Hahn condition for unlike nuclear spins in a double rotating frame \cite{H62}. 

 \begin{figure}[h]
  \centering
\includegraphics[width=0.8 \linewidth]{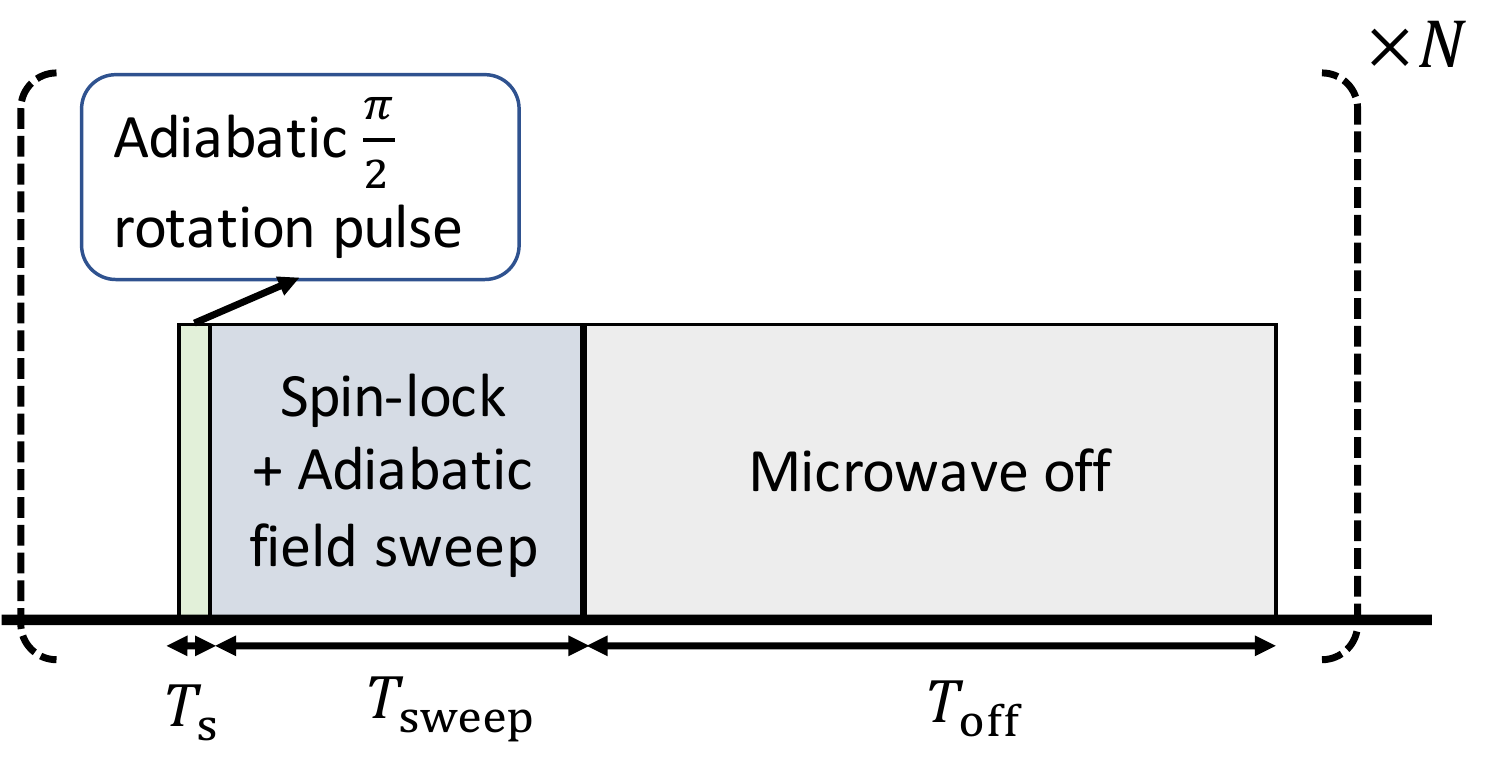}
  \caption[jsdjh]{ The schematic time line of adiabatic-NOVEL. It consists of an adiabatic rotation pulse with length $T_{s}$ followed by an adiabatic sweep of the Rabi field for a period of $T_{\text{sweep}}$. Next, the microwave field is off for some time that is long enough in comparison with the electron thermal relaxation time, initializing the electron' state. The above steps are repeated $N$ times such that the total DNP time is still less than the nuclear spin relaxation time.}
  \label{Fig_Schematic_Exp}
  \end{figure}

In one hand, it is desirable to implement NOVEL at high field in order to start with a highly polarized electron. On the other hand, according to Eq.\ref{Eq_HH}, it brings a technical challenge for engineering a device that is capable of generating very strong microwave fields at large frequencies (e.g. 10s of GHz). Furthermore, NOVEL is sensitive to the energy matching condition, as shown later in this report. In particular, when there is an unknown local magnetic field at the electron site that is comparable with the Rabi field, the polarization enhancement is reduced considerably. The extra local magnetic field could be due to field in-homogeneity or g-anisotropy of the lattice and etc. This sensitivity of NOVEL to the energy matching condition makes this DNP technique unsuitable candidate for nano-MRI applications, where field gradients are the essential ingredient of the imaging.

\begin{figure*}
  \includegraphics[width=\textwidth,height=7cm]{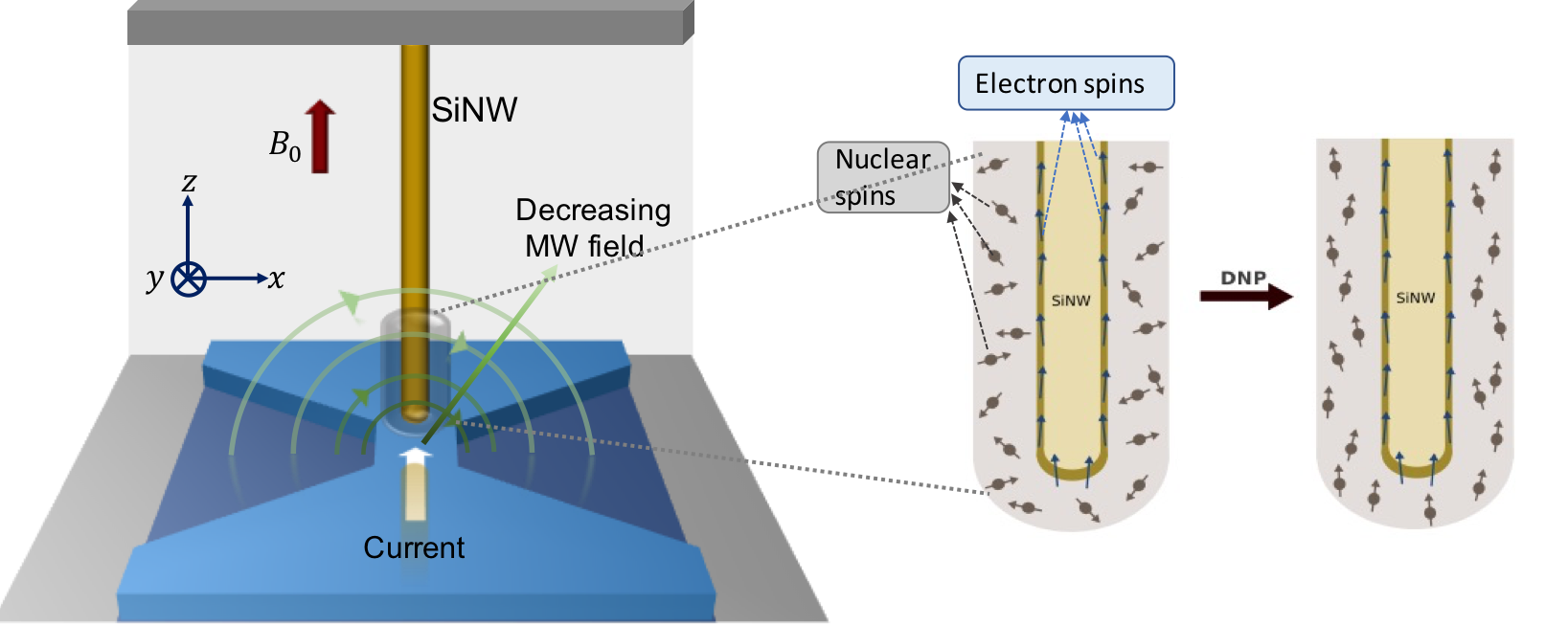}
  \caption{ A schematic representation of the MRFM imaging device. The SiNW is located right at the top of the current focusing spot of the metallic CFFGS. The white thick arrow represents the direction of electric current flow and the green curves represents contours of constant Rabi frequency near the tip of SiNW. The right figure schematically indicates the sample coated at the tip of SiNW. Once the DNP process is implemented successfully, the magnetization is transferred from the paramagnetic centers at the surface of SiNW to the protons inside the sample.}
   \label{Fig_apparatus}
\end{figure*}

Figure \ref{Fig_apparatus} schematically demonstrates a particular MRFM apparatus that was designed and engineered by \cite{N12}. It consists of a silicon nano-wire, which is an ultra-sensitive mechanical oscillator, and a metallic constriction, which is a current focusing field gradient source (CFFGS). The metallic constriction is capable of generating large time-dependent field gradients ($\sim10^6$ T/m) over a large frequency bandwidth (DC-GHz). The variation of the field is done with high precision ($0.05$ T) which enables rapid spin manipulation and control \cite{N12, N13}.  A fixed current passing through the CFFG results in a wide distribution of Rabi frequencies (e.g. 30 MHz-70 MHz) across the tip of the nano-wire (50 nm length). Consequently, at a fixed current, only a fraction of the spin ensemble can be in energy contact with their nearby electrons. Therefore, if we consider implementing the conventional NOVEL with this device, we should expect a poor efficiency because majority of spins do not effectively contribute to the DNP process. Here, we propose to apply a time-varying current instead. 

We consider an adiabatic version of NOVEL. It consists of an adiabatic $\frac{\pi}{2}$ rotation pulse with length $T_{s}$ followed by an amplitude sweep of Rabi field for $T_{\text{sweep}}$. The microwave field is then turned off for a period $T_{\text{off}}$ (Figure \ref{Fig_Schematic_Exp}) and the whole process is repeated $N$ times to optimize the DNP enhancement. For the analysis, it is convenient to treat the proposed DNP method as a two-phase process: the \textit{cross polarization} and the \textit{spin diffusion}. In the cross polarization phase \cite{ME79}, the electron's state is \textit{spin locked} to the microwave field's direction and the field's amplitude varies as $\omega_{1e}(t) \in \{ \omega_{0n} - \Delta \omega\ ,\  \omega_{0n}+\Delta \omega\}$. During the sweep time, $T_{\text{sweep}}$, the electron polarization is transferred to a group of coupled nuclear spins, whom we refer to as the \textit{nuclear cloud}. In principle, one can design various ways of amplitude modulation but we choose the linear sweep to explain the basic principles. In particular, we consider $\omega_{1e}(t) = \omega_{0n} - \Delta \omega\ (2 t/T_{\text{sweep}}-1)$, where  the electron-nuclear cross polarization occurs near $t^{*}=\frac{T_{\text{sweep}}}{2}$ when the electron Rabi frequency is on resonance with the nuclear larmor frequency. In the spin diffusion phase \cite{D08}, the transferred polarization to the nuclear cloud diffuses across the sample by means of the homonuclear dipole-dipole interaction \cite{S85, D08, Z98,Ho11,Ram08}. After the spin locking period, the microwave field is off for $T_{\text{off}}$ that needs to be much longer than the electron relaxation time, i.e., $T_{\text{off}}\gg T_{1e}$. During this period, the electron's state is rest to the thermally polarized state while the nuclear cloud polarization diffuses to the bulk. Note that the spin diffusion occurs at all times during the experiment. But, the diffusion rates are often much slower than the cross polarization rate and it is a fair approximation to neglect the diffusion during the spin locking period. We repeat the above process as much as possible with the constraint that $N(T_{s }+ T_{\text{sweep}} + T_{\text{off}})\ll T_{1n}$. This ensure us that the nuclear relaxation during the DNP process is not significant and the polarized ensemble is preserved for further imaging process.

We emphasize that the application of adiabatic-NOVEL is not limited to the cases with field in-homogeneity. As shown later, even under ideal condition when there is no field distribution, the adiabatic-NOVEL can achieve maximum polarization enhancement as opposed to the conventional NOVEL.

  For implementation of adiabatic-NOVEL via MRFM imaging device, one should note that the Rabi frequency is a function of both the electron's location and time (i.e., $\omega_{1e}(z, t)$). Therefore, in order to cover the Rabi field dispersion, one should choose $\Delta \omega$ wide enough so that all electrons across the sample have the chance to come in energy contact with their coupled nuclear spin. In other words, they all effectively contribute to the DNP process, although, the exact cross polarization time is different for individual electrons. Moreover, with the above MRFM device, it is not feasible to rotate all the electrons across the sample by a single hard pulse due to the variation in $B_{1}$. Nevertheless, the authors of \cite{GA01} have shown that an adiabatic sweep of both phase and the amplitude of the microwave field can rotate all electrons to the X-Y plane. These adiabatic pulses can be implemented very fast (a few ns) as shown in Appendix \ref{Sec_AHP}.

For the proof-of-principle experiment, we consider a water sample with the thickness of 50 nm that covers 50 nm of the SiNW length. This spin ensemble contains $N_{n}=4\times 10^{7}$ protons that is expected to exhibit a very long relaxation time at 300 mK. Regarding the polarizing agent, there are a few candidates that can be considered. One may consider injecting paramagnetic centers into the the imaging solution. This benefits us in the sense that we have control over the electron density but it may introduce some limitations for the subsequent imaging procedure. For instance, the paramagnetic impurities can shorten the proton coherence time and/or generate a considerable local magnetic field that is an undesired field inhomogeneity across the sample. Both of these may introduce extra challenges into the imaging protocol. Alternatively, one can take advantage of the paramagnetic electron spins at the interface of Si and SiO$_{2}$ of the SiNW to play the role of the polarizing agents \cite{B89, S07}. These electrons are indeed dangling bonds, also known as Pb centers, that are produced as a result of lattice constant mismatch at the interface of Si/SiO$_{2}$. Since these electrons are not inside the sample, we are not concerned about their influence on the imaging process, but, we have less control over their density and less certainty about their  location or depth relative to the surface. One trade off between the above possibilities is to inject a layer of paramagnetic centers at the surface of SiNW and then coat it with the sample solution. This way, the polarizing agent are kept outside the target sample and their density is controllable. The latter is what we consider for the rest of the analysis. 
  

\section{Theory}
\label{Sec_theory}
  In this section, we provide a theoretical description of the spin dynamics during the cross polarization phase. For the numerical estimation we consider cryogenic temperature ($T=300$ mK) and a relatively strong static field ($B_{0}= 1.2\ T$). Under this condition, the electrons are $98\%$ polarized and the protons are $0.3 \%$ polarized. Thus, it is a good approximation to consider the initial state as $\rho'_{0}= |\uparrow\rangle \langle \uparrow|_{Z} \otimes \frac{\mathbb{1}^{\otimes k}}{2^k}$ which transforms to $\rho_{0}= |\uparrow\rangle \langle \uparrow|_{X} \otimes \frac{\mathbb{1}^{\otimes k}}{2^k}$ after applying a $\pi/2$ rotation pulse. At this field, the corresponding electron and nuclear larmor frequencies are $\omega_{0e} = 33$ GHz and $\omega_{0n} = 51$ MHz respectively. The state of-the art metallic constriction that is engineered for sensitive nano-scale imaging is capable of generating very strong Rabi field (such as 50 MHz) at large frequencies (such as 33 GHz), and so, is ideal for implementing the adiabatic-NOVEL. 
 \subsection{Interaction Model}
Consider a system of an unpaired electron, $S=\frac{1}{2}$, interacting with one nuclear spin, $I=\frac{1}{2}$. In the presence of a static magnetic field, $B_{0}$, a microwave field with amplitude $2B_{1}$ and frequency $\omega_{c}$, and in the absence of any chemical shift anisotropy and/or $g$ anisotropy, the spin Hamiltonian of an electron-nuclear coupled system in the lab frame is
  \begin{eqnarray}
 H_{\text{lab}}&=& \gamma_{e} B_{0} \ (S_{z} \otimes \mathbb{1}) + \gamma_{n} B_{0}\ (\mathbb{1} \otimes  \ I_{z})\\ \nonumber
 &+& 2\gamma_{e} B_{1} \ \cos\omega_{c} t \  (S_{x} \otimes \mathbb{1}) + 2 \gamma_{n} B_{1} \ \cos\omega_{c} t \  (\mathbb{1} \otimes  \ I_{x}) \\ \nonumber
 &+& \frac{\mu_{0} \gamma_{e} \gamma_{n} \hbar}{ 4\pi} \ \frac{1}{|\vec{r}|^3} \ \left[ 3( \vec{S}. \hat{r})( \vec{I}. \hat{r}) - \vec{S}. \vec{I} \right] + \ \vec{S}. \overleftrightarrow{F}. \vec{I}, \\ \nonumber 
 \end{eqnarray}
where $\vec{r}= (R, \theta, \varphi)$ is the inner distance vector between the electron and the nuclear spin written in the Zeeman frame, $\gamma_{e}$ (or $\gamma_{n}$) is the electron (or the nuclear) gyromagnetic ratio, $\mu_{0}$ is the vacuum permeability and $\hbar=1$. The first four terms of $ H_{\text{lab}}$ are due to the interaction between the spin system and the external magnetic fields and the last two terms is due to hyperfine interaction which consists of a dipole-dipole interaction and a Fermi contact term \cite{S96}. At relatively large static field, the electron Zeeman interaction is the dominant term in the Hamiltonian. Therefore, it is valid to do a \textit{pseudo-secular} approximation where only those terms of $H_{\text{lab}}$ are considered that commute with $S_{z} \otimes \mathbb{1}$. Furthermore, after rotating wave approximation \cite{WY07}, the Hamiltonian in the rotating frame of the pulse becomes
 \begin{eqnarray}\nonumber
 \label{Eq_Ham_Rot}
 H_{\text{rot}}  
 &\approx& \left( \delta \omega \ S_{z} + \omega_{1e}\  S_{x} \right) \otimes \mathbb{1}+ \omega_{0n}\ (\mathbb{1} \otimes \ I_{z}) \\ \nonumber
 &+&    \ S_{z} \otimes \left(A \ (\cos \varphi \ I_{x}+ \sin \varphi \ I_{y}) + C \ I_{z}\right),\\ 
\end{eqnarray}
 where $\delta \omega = \omega_{0e} - \omega_{c} $ is the Zeeman off-resonance with $\omega_{0e} = \gamma_{e} \ B_{0}$ and $  \omega_{1e} =  \gamma_{e} B_{1}$ being the electron larmor and Rabi frequency respectively. Here, $A$ and $C$ are the \textit{hyperfine parameters} given by
\begin{eqnarray}\\ \nonumber
 && A =  \frac{3}{2} \ \frac{\mu_{0} \gamma_{e} \gamma_{n} \hbar}{ 4\pi R^3} \ \sin 2 \theta , \\ \nonumber
  && C =  \frac{\mu_{0} \gamma_{e} \gamma_{n} \hbar}{ 4\pi R^3} \ (3 \cos\theta^2 -1) + F_{zz}. \\ \nonumber
 \end{eqnarray}
 Since the $S_{z} \otimes I_{x}$ or the $S_{z} \otimes I_{y}$ term of the hyperfine Hamiltonian induce transition between the Zeeman energy levels of the nuclear spin, we refer to $A$ as the \textit{energy mixing parameter}. Similarly, we refer to $C$ as the \textit{energy shifting parameter} because the $S_{z} \otimes I_{z}$ term shifts the energy of the nuclear spin. {\color{black}Note that the hyperfine coupling strength scales with the inverse cubic of the electron-nuclear distance.}

\subsection{Energy Mixing}
In this section, we study the contribution of the energy mixing parameter into the cross polarization phase of the adiabatic-NOVEL. In the next section, the contribution of the energy shifting parameter is also included either as a perturbing Hamiltonian or as a strong local magnetic field.

We expand the Hamiltonian in Eq.\ref{Eq_Ham_Rot} in $\{|\uparrow\rangle_{X} , |\downarrow\rangle_{X} \} \otimes \{| \Uparrow\rangle_{Z} , |\Downarrow\rangle_{Z} \}$ basis to obtain
\begin{eqnarray}
\label{Eq_Ham_simple}
H_{0} &=&\omega_{1e}\  S_{x} \otimes \mathbb{1} + \omega_{0n}\  \mathbb{1} \otimes I_{z} + A\ S_{z} \otimes I_{x}, \\ \nonumber
&=& \frac{1}{2}\left[ \begin{array}{c |c  c| c}
\omega_{1e}+ \omega_{0n} & 0 & 0 & \frac{A}{2}\\ 
\hline
0& \omega_{1e}- \omega_{0n} & \frac{A}{2} & 0 \\ 
0&  \frac{A}{2} & - \omega_{1e}+ \omega_{0n}& 0 \\
\hline 
\frac{A}{2}& 0 & 0 & -\omega_{1e}- \omega_{0n} \\ 
\end{array}\right].
\end{eqnarray}
Here, we consider an on resonance microwave irradiation, i.e., $\delta \omega =0$, and without loss of generality, we choose a coordinate that $\varphi=0$.

\begin{figure}[h]
  \centering
\includegraphics[width=0.4 \linewidth]{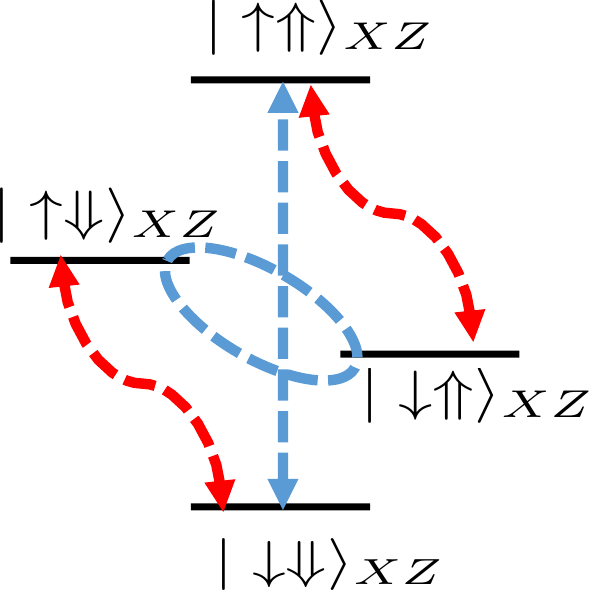}
  \caption[jsdjh]{Energy diagram of an electron-nuclear coupled system. The dashed blue lines indicate the transitions induced by energy mixing parameter, $A$, and the dashed red lines indicate the transitions induced by energy shifting parameter, $C$.}
  \label{Fig_Energy_Diagram}
  \end{figure}
  
  In this basis, one can decompose the Hilbert space as a direct sum of two subspaces: The \textit{Double Quantum} subspace (DQ) and the \textit{Zero Quantum} subspace (ZQ), 
\begin{eqnarray}
\mathcal{H}&=& \mathcal{H}_{DQ} \oplus \mathcal{H}_{ZQ},\\ \nonumber
\mathcal{H}_{DQ} &=& \text{Span} \{ | \uparrow \Uparrow \rangle _{XZ}, | \downarrow \Downarrow \rangle _{XZ}\},\\ \nonumber
\mathcal{H}_{ZQ} &=& \text{Span} \{ | \uparrow \Downarrow \rangle _{XZ}, | \downarrow \Uparrow \rangle _{XZ}\}.\\ \nonumber
\end{eqnarray}
 Each of the above subspaces can be considered as a \textit{pseudo-spin half} with associated Pauli Operators as,
 \begin{eqnarray}\nonumber
 \label{Eq_Pauli}
 \sigma_{z}^{DQ} &:=&  | \uparrow \Uparrow \rangle \langle \uparrow \Uparrow | - | \downarrow \Downarrow \rangle \langle \downarrow \Downarrow |, \\ \nonumber
   \sigma_{X}^{DQ} &:=&  | \uparrow \Uparrow \rangle \langle \downarrow \Downarrow | + | \downarrow \Downarrow \rangle \langle \uparrow \Uparrow | ,\\ \nonumber\\ \nonumber
  \sigma_{z}^{ZQ} &:=&  | \uparrow \Downarrow \rangle \langle  \uparrow \Downarrow | - | \downarrow \Uparrow \rangle \langle \downarrow \Uparrow |,  \\ \nonumber
    \sigma_{X}^{ZQ}& :=&  | \uparrow \Downarrow \rangle \langle \downarrow \Uparrow | + | \downarrow \Uparrow \rangle \langle \uparrow \Downarrow |.  \\
 \end{eqnarray}
 
Given the above definitions, one can re-write Eq. \ref{Eq_Ham_simple} as
 \begin{eqnarray}
 H_{0}&=& H^{DQ} \oplus H^{ZQ},\\ \nonumber
 &=& \frac{\omega_{1e}+ \omega_{0n} }{2} \ \sigma_{z}^{DQ} + \frac{A}{4} \ \sigma_{X}^{DQ}  \\ \nonumber 
&\oplus& \frac{\omega_{1e}- \omega_{0n} }{2} \ \sigma_{z}^{ZQ} + \frac{A}{4} \ \sigma_{X}^{ZQ} . \\ \nonumber
 \end{eqnarray}
 %

Since $H_{0}$ is block diagonal, one can diagonalize each subspace separately. We denote the eigenbasis of the $DQ$ subspace with $\{|\Psi_{+}\rangle\ ,\ | \Psi_{-}\rangle\}$ and the eigenbasis of the $ZQ$ subspace with $\{|\Phi_{+}\rangle\ ,\ | \Phi_{-}\rangle\}$ with the corresponding eigenenergies
   \begin{eqnarray}\nonumber
  \label{Eq_Omega}
\Omega_{_{DQ}} &:=&\pm  \frac{1}{2} \sqrt{ ( \omega_{1e} + \omega_{0n} )^2 + (\frac{A}{2})^2},  \\ \nonumber
\Omega_{_{ZQ}} &:=&  \pm  \frac{1}{2} \sqrt{ ( \omega_{1e} - \omega_{0n} )^2 + (\frac{A}{2})^2}.  \\
\end{eqnarray}

Similarly, one can write the initial density matrix in terms of the above defined Pauli operators. After a $\frac{\pi}{2}$ rotation, the state of a polarized electron interacting with an unpolarized nuclear spin is given by $\rho_{0}= |\uparrow\rangle \langle  \uparrow|_{X} \otimes \frac{\mathbb{1}}{2}$, which can be expanded as  
\begin{equation}
\label{Eq_ini}
 \rho_{0}= \frac{1}{2}\  ( \frac{\mathbb{1}^{DQ} + \sigma_{Z}^{DQ}} {2} + \frac{\mathbb{1}^{ZQ} + \sigma_{Z}^{ZQ}} {2})
\end{equation}
Thus, the initial state of interest has $50\%$ polarization in the $DQ$ subspace and $50\%$ polarization in the $ZQ$ subspace and each subspace evolves according to \textit{its own} Hamiltonian ($H^{DQ}$ or $H^{ZQ}$). Moreover, 
 \begin{eqnarray}
 \label{Eq_Sigmas}
 \sigma_{x} \otimes \mathbb{1}&=& \sigma_{Z}^{DQ} + \sigma_{Z}^{ZQ},\\ \nonumber
 \mathbb{1} \otimes \sigma_{z}&=& \sigma_{Z}^{DQ} - \sigma_{Z}^{ZQ}. \\ \nonumber
 \end{eqnarray}
 The above relations imply that the electron polarization in the rotating frame is a sum of the polarization in the $DQ$ subspace and that of the $ZQ$ subspace, whereas, the nuclear polarization is the difference between the two.
 
   In the case of conventional NOVEL, the electron Rabi frequency is constant, $\omega_{1e} \sim \omega_{0n}$, and so, the evolution operator is given by
 \begin{eqnarray}
 U&=& \exp[- i t \  H_{0}] \\ \nonumber
 &=& \cos[\frac{\Omega_{_{DQ}} t}{2}] \ \mathbb{1}^{DQ} - i \sin[ \frac{\Omega_{_{DQ}} t}{2}] ( \cos\psi \ \sigma_{Z}^{DQ} + \sin\psi \ \sigma_{X}^{DQ}) \\ \nonumber 
 & \oplus & \cos[\frac{\Omega_{_{ZQ}} t}{2}]\ \mathbb{1}^{ZQ} - i \sin[ \frac{\Omega_{_{ZQ}} t}{2}] ( \cos\phi \ \sigma_{Z}^{ZQ} + \sin\phi \ \sigma_{X}^{ZQ}).
 \end{eqnarray}
Here, the $\psi= \tan^{-1}[ \frac{A/2}{ \omega_{1e} + \omega_{0n}}]$ and the $\phi= \tan^{-1}[\frac{A/2}{ \omega_{1e} + \omega_{0n}}]$ are the quantization angles in the DQ and the ZQ subspace respectively. Given the initial state, $\rho_{0}$ and the unitary, $U$, the nuclear magnetization evolves as

\begin{eqnarray}
\label{Eq_net_enhancement}
M_{n}(t) &=& \frac{\hbar\ \gamma_{n}}{2}\ Tr[ (\mathbb{1} \otimes \sigma_{z}) ( U \ \rho_{0} \ U^{\dagger})] \\ \nonumber
&=&\frac{\hbar\ \gamma_{n}}{2}( \left[ \cos^2 \psi + \cos \Omega_{_{DQ}} t\  \sin^{2} \psi \right]\\ \nonumber
 &&\hskip 0.5 cm - \left[ \cos^2 \phi + \cos \Omega_{_{ZQ}} t \ \sin^{2} \phi \right] ).
\end{eqnarray}

Consider an \textit{intermediate} hyperfine coupling limit where $(\omega_{1e}- \omega_{0n})\ll\frac{ A}{2} \ll (\omega_{1e}+ \omega_{0n})$, and so, $\psi\sim 0$ and $\phi \sim \frac{\pi}{2}$. In this limit, the electron-nuclear interaction is too \textit{weak} to mix the population of energy levels in the double quantum subspace. As a result, the $50\%$ polarization of the $DQ$ subspace in Eq. \ref{Eq_net_enhancement}) remains steady and does not contribute to the polarization transfer process. But, the coupling is \textit{strong} enough to induce transitions between the energy levels in $SQ$ subspace. Thus, the $50\%$ polarization of the $ZQ$ subspace experiences an oscillation between the electron and the nuclear spin. Therefore, on average and under ideal conditions, the conventional NOVEL results in a $50\%$ enhancement of the nuclear polarization. We consider the intermediate coupling limit because the MRFM apparatus can generate $\omega_{1e}+ \omega_{0n} \sim 100$ MHz which is much stronger than the hyperfine coupling of a typical proton at a distance of $ 2 \ A^{0}$ from the electron ($\frac{A}{2} \sim 5$ MHz).

In the case of adiabatic-NOVEL, the amplitude of the Rabi field is not constant in time, but rather varies linearly as $ \omega_{1e}(t)- \omega_{0n}= \Delta \omega\ (2\  t/ T_{\text{Sweep}}-1)$. In fact, the electron is spin locked to a microwave field with a fixed direction but with a variable amplitude. Like before, in the limit of intermediate hyperfine coupling, the contribution of the $DQ$ subspace to the polarization transfer process is negligible. Thus, we elaborate on the spin dynamics in the the ZQ subspace. Consider the $\text{Span} \{|0\rangle:= |\uparrow \Downarrow\rangle,|1\rangle:= |\downarrow \Uparrow\rangle \}$ as a \textit{pseudo-spin half} system. The corresponding ZQ Hamiltonian is 
\begin{equation}
\label{Eq_Landau}
H^{\text{ZQ}}= \frac{\omega_{1e}(t) - \omega_{0n}}{2}\ \sigma_{z}^{ZQ} + \frac{A}{4}\  \sigma_{x}^{ZQ}.
\end{equation}
 \begin{figure}[h]
\includegraphics[width=0.8\linewidth]{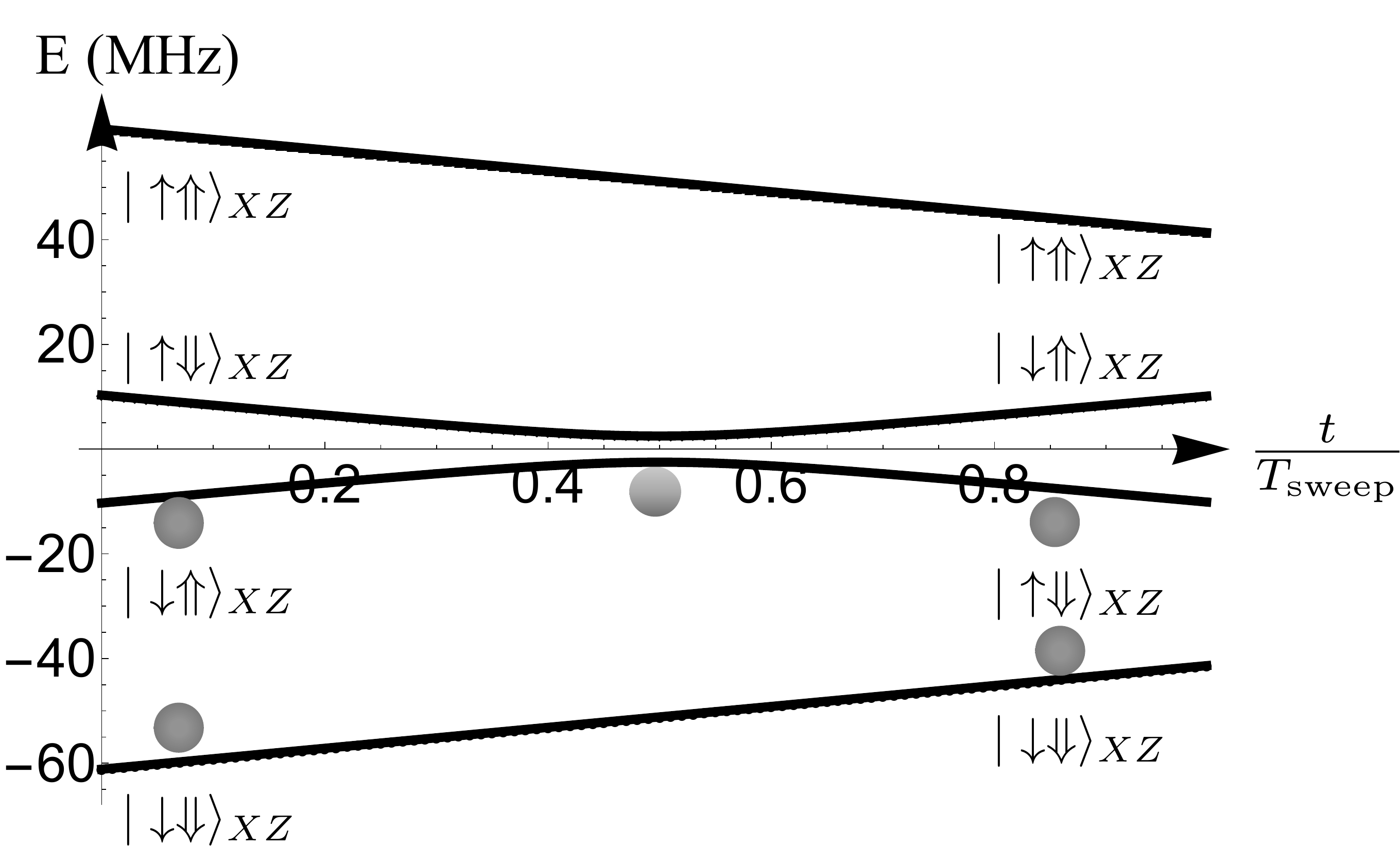}
 \caption{\label{Fig_Engen_Energy}The eigenenergies of an electron-nuclear system are shown where the amplitude of the electron Rabi field is adiabatically swept through the nuclear larmor frequency. If the sweeping rate is slow enough in comparison with the electron-nuclear hyperfine coupling, one can convert the population from the $|\uparrow \Downarrow\rangle$ state to the $ |\downarrow \Uparrow\rangle$ state. }
 \end{figure}

\begin{figure}[t]
\includegraphics[width=0.9\linewidth]{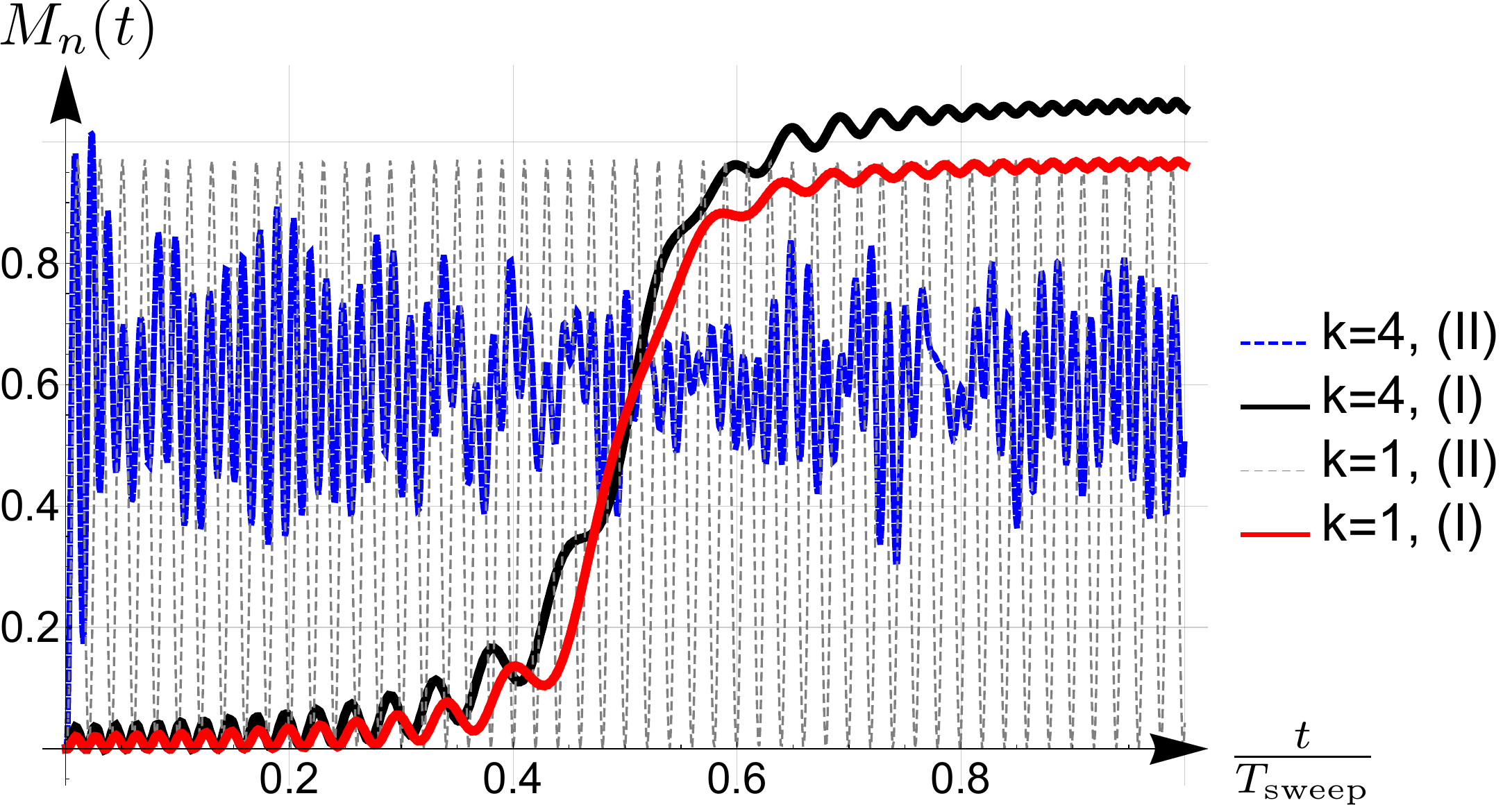}
 \caption{\label{Fig_cloud}{ Compares the acquired nuclear polarization through the adiabatic-NOVEL (I) versus the conventional NOVEL (II) for two cloud size of $k=1$ and $k=4$. Two water molecules are randomly chosen within a distance of $2-4 \ A^{0}$ from the surface of paramagnetic centers. The corresponding hyperfine parameters are $A\in \{ 4.7\ \text{MHz}, 5.6\  \text{MHz} ,-59\  \text{KHz}, 1.83343 \ \text{MHz} \}$ and $C\in\{ -1.56\ \text{MHz}, 8.8\  \text{MHz} ,-29  \ \text{KHz}, -17\  \text{KHz} \}$. The proton-proton dipolar coupling is also included.}}
 \end{figure}

The $ZQ$ subspace of an electron-nuclear system resembles the Landau-Zener effect for a single spin \cite{Zener32}.  According to this effect, if we start from a state $|\uparrow \Downarrow\rangle$ and evolve it, using Eq. \ref{Eq_Landau}, the probability of it remaining in that state is:
\begin{equation}
\label{Eq_Land_Prob}
P=e^{-2\pi \gamma} \hskip 1 cm \text{where} \hskip 1 cm \gamma= \frac{1}{T_{\text{Sweep}}} \frac{|A/2|^2}{ \frac{d\Delta\omega_{1}}{ds}},
\end{equation}
where $s= \frac{t}{T_{\text{sweep}}}$ is a dimensionless parameter with $T_{\text{sweep}}$ being the total sweeping time. From the above relation we see that when the sweeping rate is very slow in comparison with the hyperfine mixing parameter, $\frac{A}{2}$, one can make this transition very negligible and adiabatically convert all the population in the $  |\uparrow \Downarrow\rangle$ state to the $|\downarrow \Uparrow\rangle$ state. This is a great advantage over the conventional NOVEL where the polarization coherently oscillates between the electron and the nuclear spin. There, the average polarization enhancement is reduced at least by a factor of $\frac{1}{2}$. The high efficiency of adiabatic-NOVEL stems from the fact that one can adiabatically transfer all the electron polarization to the coupled nuclear spin. Figure \ref{Fig_cloud} compares the performance of these two NOVEL techniques (constant vs time-dependent Rabi field) for two cloud size of $k=1$ and $k=4$ coupled nuclear spins. 

\begin{figure}[h]
\includegraphics[width=0.8\linewidth]{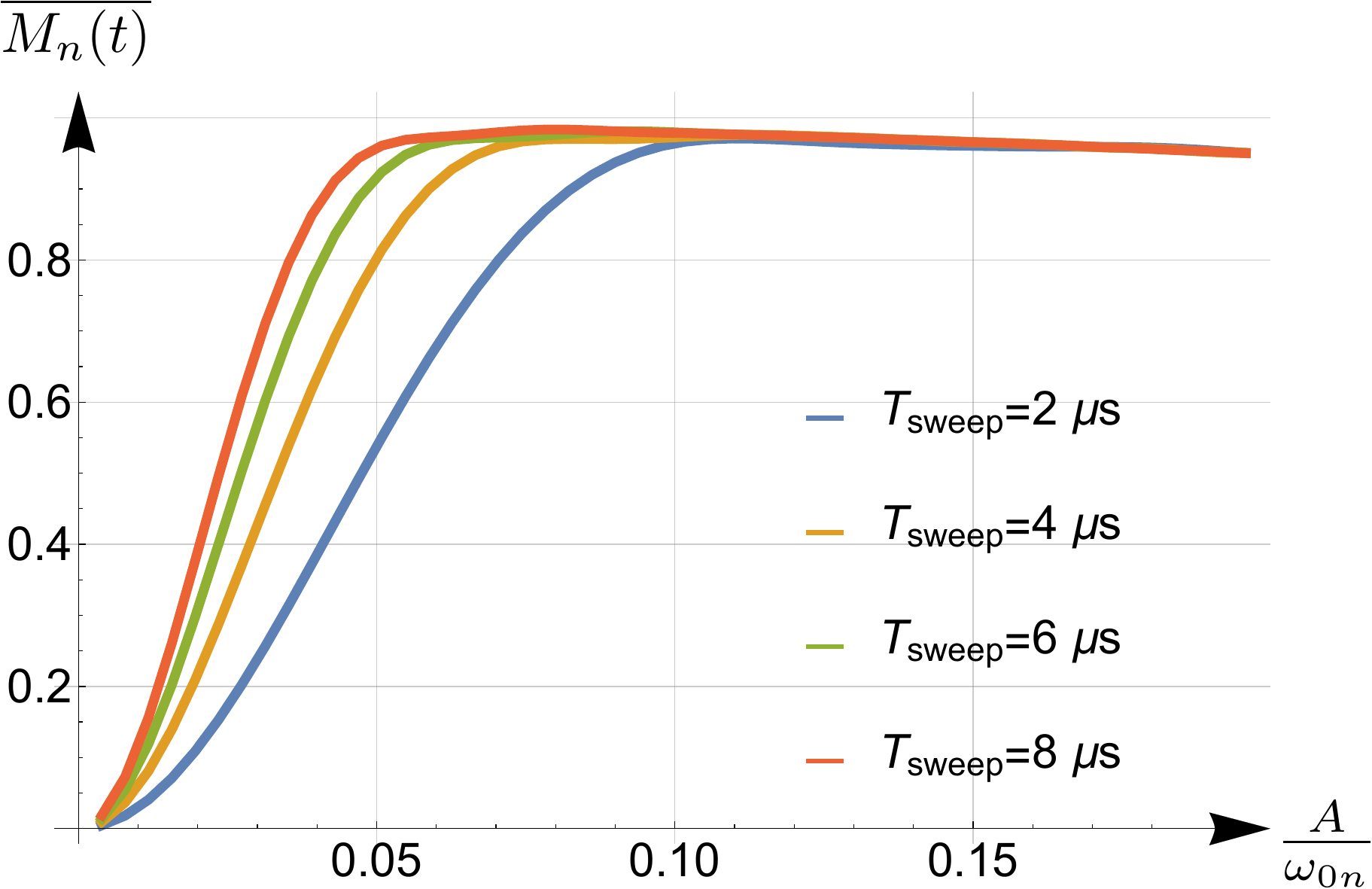}
 \caption{\label{Fig_Final-Polz-LSweep1}The final nuclear polarization enhancement via the adiabatic-NOVEL technique is plotted as a function of electron-nuclear hyperfine mixing parameter. These results show that a particular proton with a certain coupling can acquire different polarization enhancement depending on the sweeping rate of the microwave field. $\Delta \omega = 20$ MHz and $\omega_{0n} = 50$ MHz were chosen for the simulation.   }
 \end{figure}

For realization of adiabatic-NOVEL, one should choose the sweeping rate wisely. At a certain sweeping rate, $\Delta \omega /T_{\text{Sweep}}$, different nuclear spins with different coupling strengths acquire different polarization enhancements, as the simulation in Fig.\ref{Fig_Final-Polz-LSweep1} shows. For the MRFM application, the choice of the sweeping bandwidth, $\Delta \omega$ is constrained by the the width of the Rabi field distribution across the sample. Thus,  according to Eq. \ref{Eq_Land_Prob}, the only control parameter is $T_{\text{sweep}}$. One can choose a very long sweeping time so that even the weakly coupled nuclear spins exchange their polarization with the electron effectively. However, the choice of sweeping time is not arbitrary due to a few constraints. First, the sweeping rate must be fast enough so that the electron relaxation during the spin locking is negligible. Second, it must be slow enough so that at least a few nearby coupled protons contribute to the cross polarization. Therefore,  $(A)^{-1}\ll T_{\text{sweep}}\ll T_{1\rho}$.

 

The final remark is that in the limit of very strong hyperfine interaction where $A \sim (\omega_{1e}+ \omega_{0n})$, the mixing of energies in the double quantum subspace is no longer forbidden. As a result, the net polarization enhancement is suppressed considerably. The reason is that according to Eq. \ref{Eq_Sigmas}, the $ZQ$ and the $DQ$ subspaces contribute to the nuclear polarization enhancement in opposite directions. Nevertheless, the MRFM imaging device of interest avoids this scenario because it operates at high field, ($\omega_{0n}\sim 50-100$ MHz) and generates strong microwave fields ($\omega_{1e} \sim 30-100$ MHz). Therefore, the protons that are at a distance of $2\ A^{0}$ or further from the electrons have a hyperfine coupling much smaller than $\omega_{0n}+ \omega_{1e}$. 


 \subsection{Energy Shifting}
So far,  the $H_{1}= C\ S_{z} \otimes I_{z}$ term of the hyperfine interaction has not been included in the analysis. In the case of a weak energy shifting interaction, where $C\ll A$ and $\omega_{0n}$ (or $\omega_{1e}$), we treat $H_{1}$ as a perturbing Hamiltonian. According to the Fermi-Golden rule, $H_{1}$ will induce transition between the states $|\Psi_{\pm}\rangle $ and $|\Phi_{\pm}\rangle $ of the unperturbed Hamiltonian, $H_{0}$, with the corresponding rates
\begin{eqnarray}
\label{Eq_transition_rates}
\Gamma_{\Phi_{\pm} \leftrightarrow \Psi_{\pm}}= \frac{C}{4} &\times &  |\frac{\sin(\psi - \phi)}{\Omega_{_{DQ}}- \Omega_{_{ZQ}}}|  \\ \nonumber
\Gamma_{\Phi_{\pm} \leftrightarrow \Psi_{\mp}}= \frac{C}{4} &\times &  |\frac{\cos(\psi - \phi)}{\Omega_{_{DQ}}+ \Omega_{_{ZQ}}}|  \\ \nonumber
\end{eqnarray}
  According to Eq.\ref{Eq_Sigmas}, the nuclear polarization is proportional to the difference of polarization between $\mathcal{H}_{DQ}$ and $\mathcal{H}_{ZQ}$ and the above induced transitions mixes the population between the two subspaces. Therefore, a non-zero value of energy shifting interaction ($C\neq 0$) acts as a \textit{polarization leakage} that reduces the net polarization enhancement. 

 In the limit of $(\omega_{1e}- \omega_{0n})\ll\frac{ A}{2} \ll (\omega_{1e}+ \omega_{0n})$, we approximate $\Gamma_{\Phi_{\pm} \leftrightarrow \Psi_{\mp}} \approx \frac{C}{\omega_{1e}+\omega_{0n} }$. Since we assumed $C\ll \omega_{0n}$ or $\omega_{1e}$, the above leakage rates are negligible. This argument is valid for both constant Rabi field and an adiabatically modulated Rabi field. In the former, the leakage rates are constant in time whereas in the latter they are time-dependent functions. In particular case of adiabatic-NOVEL, when the time varies as $t=0\rightarrow t^{*} \rightarrow T_{Sweep}$, with $t^{*}$ being the cross polarization time, the leakage rates approximately scale as $\frac{C}{2\omega_{0n} -\Delta \omega} \rightarrow \frac{C}{2\omega_{0n}} \rightarrow \frac{C}{2\omega_{0n} +\Delta \omega}$. Therefore, when the energy shifting term is small relative to the Rabi field, the polarization leakage rates are negligible for both the conventional NOVEL and the adiabatic-NOVEL. 

\begin{figure}[h]
  \centering
\includegraphics[width=0.8\linewidth]{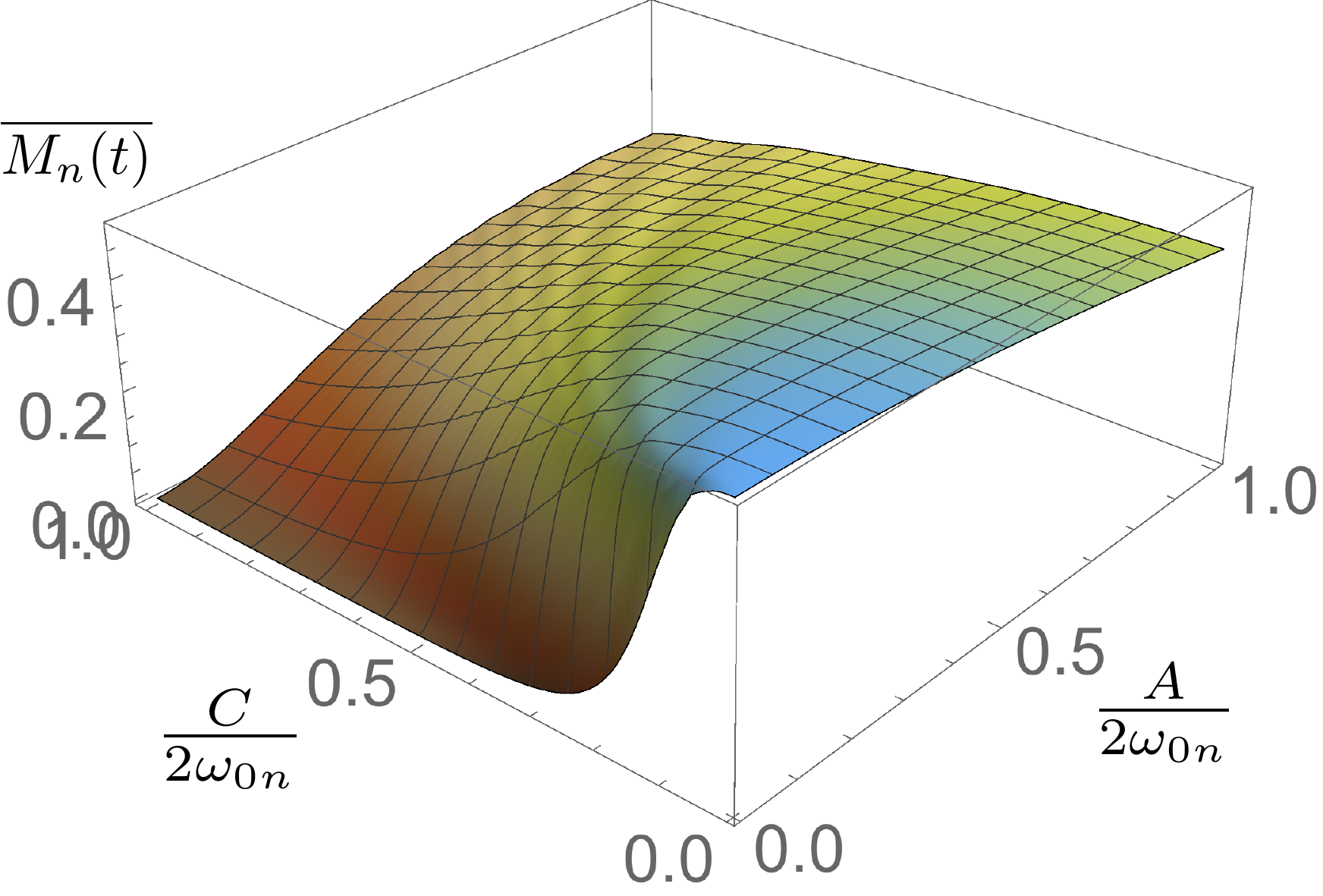}
  \caption[]{The average polarization enhancement of a nuclear spin acquired via the conventional NOVEL is plotted as a function of the relative hyperfine parameters. }
  \label{Fig_VaryAB}
\end{figure}

We numerically evaluate the net polarization enhancement acquired through conventional NOVEL for wide range of energy mixing and energy shifting parameters and plot the result in Fig.\ref{Fig_VaryAB}. It shows that this technique is very sensitive to the variation of $C$, particularly when $A \ll \omega_{0n}$. That sensitivity can be explained with the above model. In the intermediate limit of A, a small $C$ leads to a polarization leakage by scrambling the population between the ZQ and the DQ subspaces. In the limit of strong energy shifting interaction, where $C \geq \omega_{1e}$ (or $\omega_{0n}$), coupling to the nuclear spin creates a considerable local magnetic field at the electron site that is comparable to the electron Rabi frequency. In that case, the previous perturbation theory approach is not valid anymore. This strong coupling to the nearby nuclear spins can be treated as an effective magnetic filed at the electron site as,
\begin{equation}
\gamma_{e}\ \vec{\textbf{B}}\text{eff}=  \omega_{1e} \ \hat{x}\  \pm \ \frac{C}{2} \ \hat{z}
\end{equation} 
where $+$ or the $-$ sign depends on the nuclear state being $|\Uparrow\rangle$ or $|\Downarrow\rangle$. In the next secition, we study the effect of an extra local magnetic field on the DNP performance, where a strong energy shifting interaction can be considered as a particular example.

\begin{figure}[h]
\includegraphics[width=0.9\linewidth]{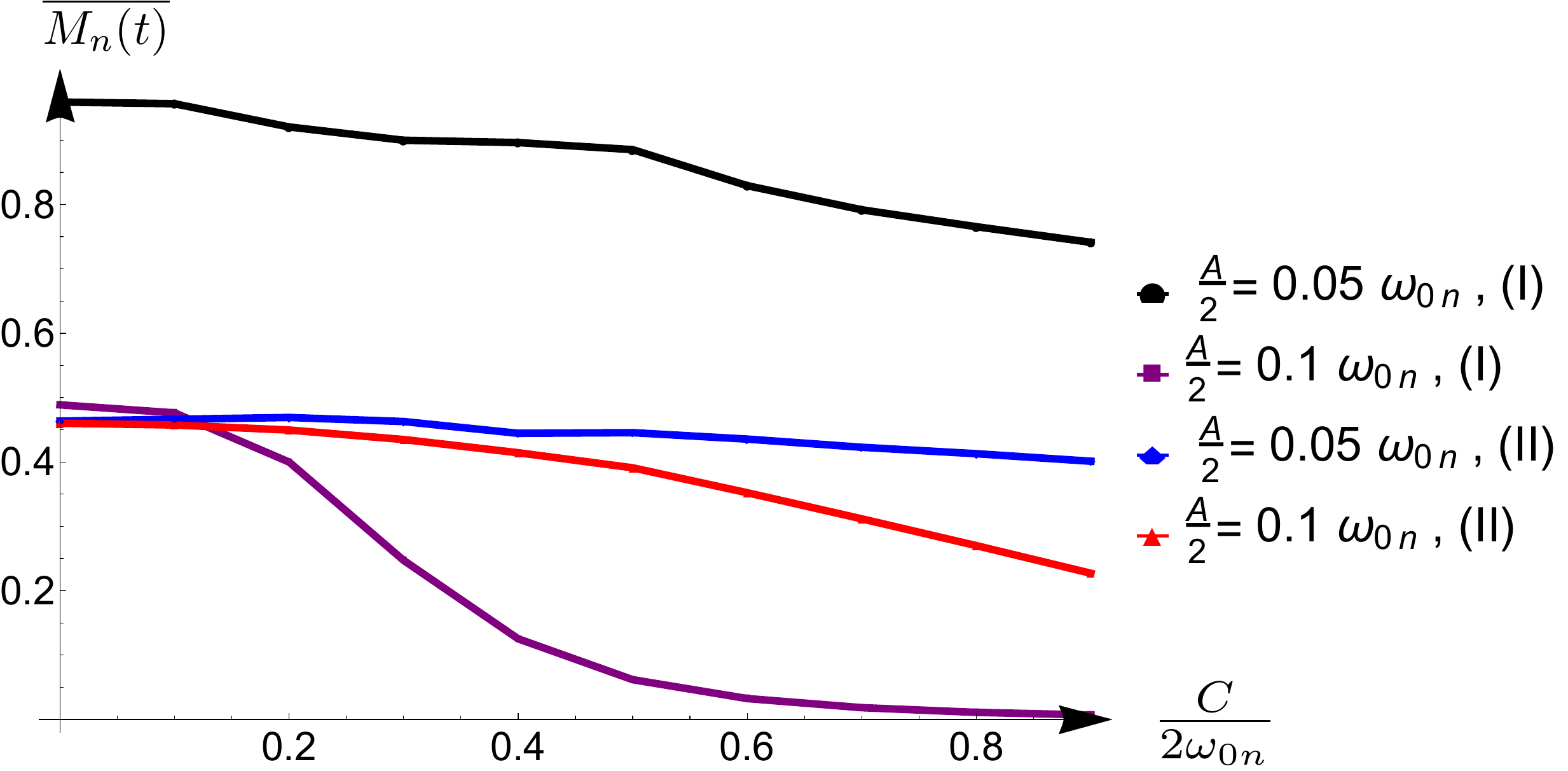}
 \caption{\label{Fig_VaryABSubfigure3} The nuclear polarization acquired via NOVEL is plotted as a function of energy shifting parameter, C. Two protons are considered, one with a relatively strong A (blue and red) and one with an intermediate energy mixing interaction (black and purple). The result compares the performance of  the adiabatic-NOVEL (I) versus the conventional NOVEL (II) in terms of efficiency and robustness. }
 \end{figure}

The goal of the simulation in Figure \ref{Fig_VaryABSubfigure3} is to compare the efficiency and robustness of the adiabatic-NOVEL with that of the conventional NOVEL. We consider a depolarized proton interacting with a fully polarized electron that is prepared along the X direction. The hyperfine mixing parameter is chosen in the intermediate regime so that the contribution of the DQ subspace is negligible (black and purple). We calculate the final polarization of the nuclear spin acquired via NOVEL and plot it for various strengths of the energy shifting term. The result shows that one can ideally achieve $100\%$ polarization transfer by means of an adiabatic amplitude modulation of the Rabi field as opposed to $50\ \%$ in case of a constant Rabi field. Furthermore, the adiabatic-NOVEL is much more robust to the variation in local magnetic field's strength. Even in the limit of strong energy mixing term (blue and red) where the efficiency of the adiabatic-NOVEL is close to that of the conventional NOVEL, the former is much more robust against undesired local interactions.

 \subsection{Robustness Against Local Magnetic Field}
 \label{Sec_robust}
In addition to the external static field, there is often an extra local magnetic field at the electron site due to various reasons such as chemical shift anisotropy, $g-$anisotropy, field in-homogeneity and interaction with other spins in the lattice. In this section, we show that the adiabatic-NOVEL is very robust against local magnetic fields in comparison with the conventional NOVEL. To explain the concept, we consider a simple example where the local interactions result in an extra term $\delta \omega_{0}\ S_{z} \otimes \mathbb{1}$ in the Hamiltonian in Eq.\ref{Eq_Ham_Rot}. That corresponds to a shift in the electron resonance frequency, $\omega_{0e}\rightarrow\omega_{0e} + \delta\omega_{0}$, but since its amount is not known and it might vary across the sample, we cannot resolve the issue by a simple shift to the microwave resonance frequency. We simplify the discussion by considering this particular local field but the same argument can be extended to a more complicated local magnetic field.

When $\delta \omega_{0}\neq 0$, in the rotating frame of the pulse, the electron is spin locked to an effective field along the $x'$ axis, that has an angle $\theta_{s}$ with the Rabi field's direction, $x$,   
\begin{eqnarray}
\hat{x'}= \cos\theta_{s} \ \hat{x} + \sin\theta _{s} \ \hat{z} \hskip 0.5 cm
\text{where} \hskip 0.5 cm
\tan\theta_{s}:= \frac{\delta \omega_{0}}{\omega_{1e}}
\end{eqnarray}
We re-write the Hamiltonian in Eq.\ref{Eq_Ham_Rot} in a basis that the electron's axis is rotated by $\theta_{s}$. i.e., in $\{|\uparrow \rangle _{x'}, |\downarrow \rangle _{x'}\} \otimes \{|\Uparrow \rangle _{Z}, |\Downarrow \rangle _{Z}\} $ basis we obtain, 

\begin{eqnarray}\nonumber
\label{Eq_H_tilde}
H_{\text{rot}} &=& \widetilde{\omega}_{\text{eff}} \ S_{x'} \otimes \mathbb{1} + \omega_{0n} \ \mathbb{1} \otimes I_{z} + A \ \cos\theta_{s} \ S_{z'} \otimes I_{x}\\ \nonumber
&+&  -A \ \sin\theta_{s} \ S_{x'} \otimes I_{x} \\ \nonumber
&+& C\ \cos\theta_{s} \ S_{z'} \otimes I_{z} -C\ \sin\theta_{s} \ S_{x'} \otimes I_{z} \\ \nonumber
&=& \widetilde{H}_{0}+ \widetilde{H}_{1}\\ \nonumber
&\text{with}& \\ 
\widetilde{\omega}_{\text{eff}} &:=& \sqrt{\delta \omega_{0}^2+\omega_{1e}^2}
\end{eqnarray}

Here, $\widetilde{H}_{0}$ contains the first three terms and its structure is similar to $H_{0}$ in Eq.\ref{Eq_Ham_simple}. Therefore, the corresponding eigenenergies are similar to Eq.\ref{Eq_Omega} except that the $\omega_{1e}$ is replaced with the $\widetilde{\omega}_{\text{eff}}$ and the $A$ with the $A \cos \theta_{s}$. The last three terms of Eq.\ref{Eq_H_tilde} are represented by $\widetilde{H}_{1}$ that is treated as a perturbation and is neglected for now but will be included later.

In case of conventional NOVEL and in the presence of an extra local field with $\delta \omega_{0}\neq 0$, the polarization exchange between the electron and the coupled nuclear spin is not optimized at the resonance condition, $\omega_{1e}= \omega_{0n}$. First, there is an energy mismatch $\Delta E = \sqrt{(\delta \omega_{0}) ^2+ \omega_{1e}^2} - \omega_{0n}$ between the electron and the nuclear spin, and second,  their effective  energy mixing interaction is now replaced as $A \rightarrow A \cos\theta_{s}$.  As a result, the polarization exchange rate is reduced at least by a factor of $\frac{\cos\theta_{s}}{\Delta E}$. This explains the result in Figure \ref{Fig_VaryAB} since a strong energy shifting parameter acts as a local magnetic field and that leads to a significant reduction in the polarization enhancement. 

\begin{figure}[h]
\includegraphics[width=0.9\linewidth]{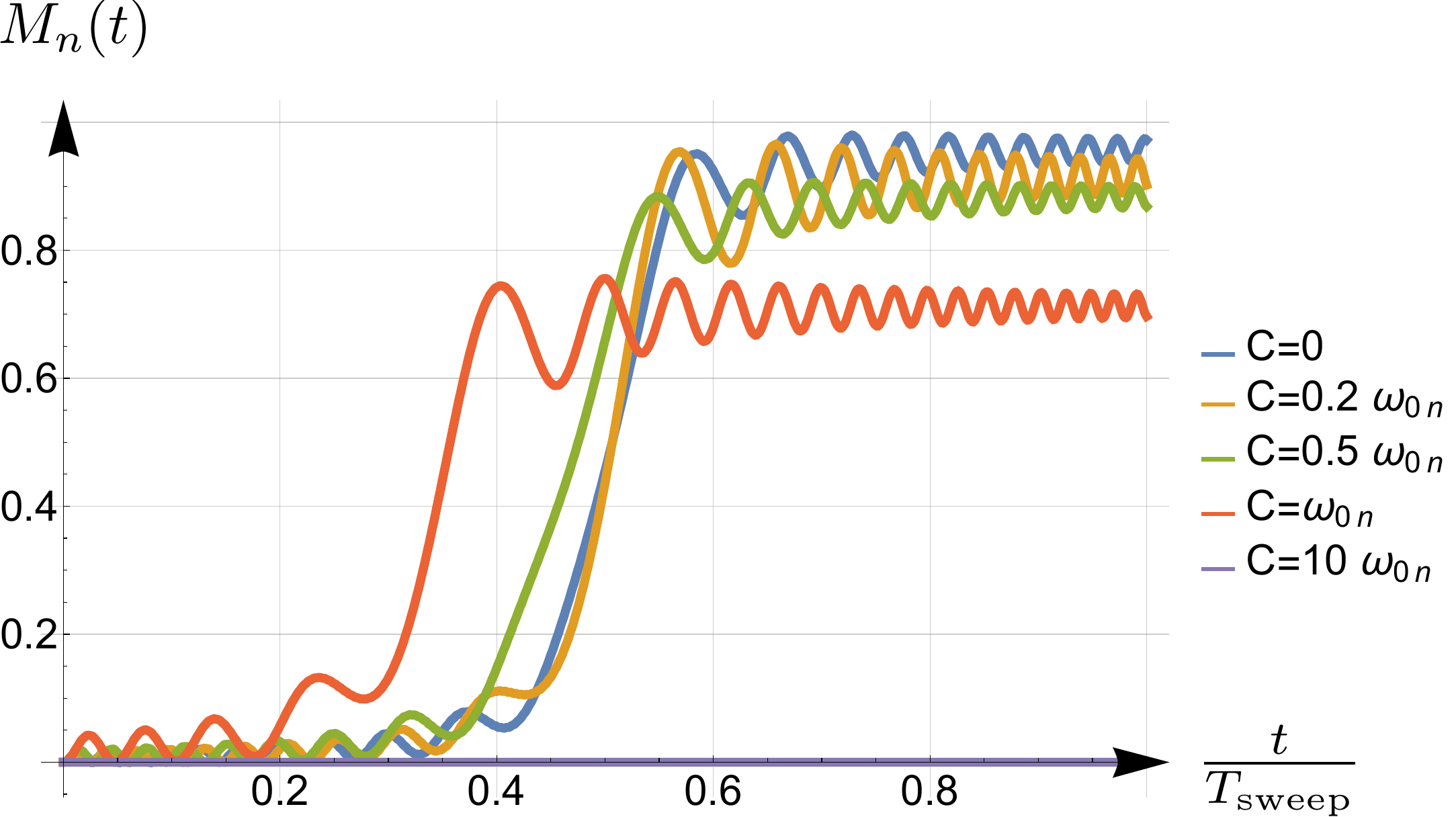}
 \caption{\label{Fig_VaryC}Nuclear polarization is plotted as a function of spin locking time in adiabatic-NOVEL for various value of the energy shifting interaction, $C$. The sweeping parameters are chosen as $\omega_{0n} \pm 20$ MHz and $T_{\text{sweep}}=2 \ \mu s$ to satisfy the adiabatic condition for a nuclear spin with $A= 0.1 \omega_{0n}$. A considerable local magnetic field leads to a shift in the cross polarization time without a dramatic reduction in the DNP enhancement.}
 \end{figure}

In case of adiabatic-NOVEL, when $\delta \omega_{0}=0$, the electron is spin locked to the microwave field that has a fixed direction $\hat{x}$ but its amplitude is varying by time. Whereas, in the presence of a considerable local magnetic field with $\delta \omega_{0} \leq \omega_{1e}$, the electron is spin locked to an effective field, $\gamma_{e} \vec{\mathbf{B}}_{\text{eff}}= \omega_{1e}(t) \ \hat{x} + \delta \omega_{0}\ \hat{z}$. Thus, both the amplitude and direction of locking field are variable in time. However, the change occurs adiabatically. Therefore, one can assume that at each instance of time, the electron \textit{instantaneous} locking axis is $\hat{x}'(t)$, that has an angle $\theta_{s}(t)$ with the microwave field's direction. Note that the \textit{electron instantaneous basis}, $\{|\uparrow \rangle _{x'(t)}, |\downarrow \rangle _{x'(t)}\}$, the effective field angle $\theta_{s}(t)$ and the effective Rabi frequency $\widetilde{\omega}_{\text{eff}}(t)$ are all time-dependent functions. This time dependency of the electron Rabi frequency implies that the electron' effective Rabi frequency can be \textit{in energy contact} with the nuclear larmor frequency at some point during the sweep. In fact, the existence of an undesired local magnetic filed is now translated to a shifting in the cross polarization time as,
\begin{equation}
\sqrt{\delta \omega_{0}^2+\omega_{1e}(t^*)^2} = \omega_{0n}
\end{equation} 
In other words, a $\delta \omega \neq 0 $ is no longer equivalent to a violation of the Hartman-Han condition but rather just a shift in the polarization crossing time. The above argument explains the numerical result presented in Fig.\ref{Fig_VaryC}, where a local field as large as $C= 0.5 \omega_{0n}$ leads to a shift in the cross polarization time of adiabatic-NOVEL with a $1\%$ reduction in the DNP enhancement. That cause of that reduction is explained in the following. 

The $\widetilde{H}_{1}$ of Eq.\ref{Eq_H_tilde} was neglected so far. Similar to the previous section, we treat this term as a perturbing Hamiltonian that leads to a polarization leakage. One can compute the transitions between the eigenstates of $\widetilde{H}_{0}$ in ZQ subspace($\{ \widetilde{\Psi}_{\pm}\}$) and that of the $DQ$ subspace ($\{ \widetilde{\Phi}_{\pm} \} $) to obtain
 
\begin{eqnarray}\nonumber
\label{Eq_transition_rates_2}
\Gamma_{\widetilde{\Phi}_{\pm} \leftrightarrow \widetilde{\Psi}_{\pm}}&=& \frac{( A \sin\theta_{s} \ \sin(\tilde{\psi} - \tilde{\phi}) + C \cos\theta_{s} \ \cos(\tilde{\psi} + \tilde{\phi}))}{4( \widetilde{\Omega}_{_{DQ}} -\widetilde{\Omega}_{_{ZQ}})}  \     \\ \nonumber
\Gamma_{\widetilde{\Phi}_{\pm} \leftrightarrow \widetilde{\Psi}_{\mp}}&=& \frac{( A \sin\theta_{s} \ \cos(\tilde{\psi} - \tilde{\phi})+ C\ \cos\theta_{s} \ \sin(\tilde{\psi} + \tilde{\phi})) }{4( \widetilde{\Omega}_{_{DQ}} +\widetilde{\Omega}_{_{ZQ}})} \ \\
\end{eqnarray}

\begin{figure}[h]
  \centering
\includegraphics[width=0.8\linewidth]{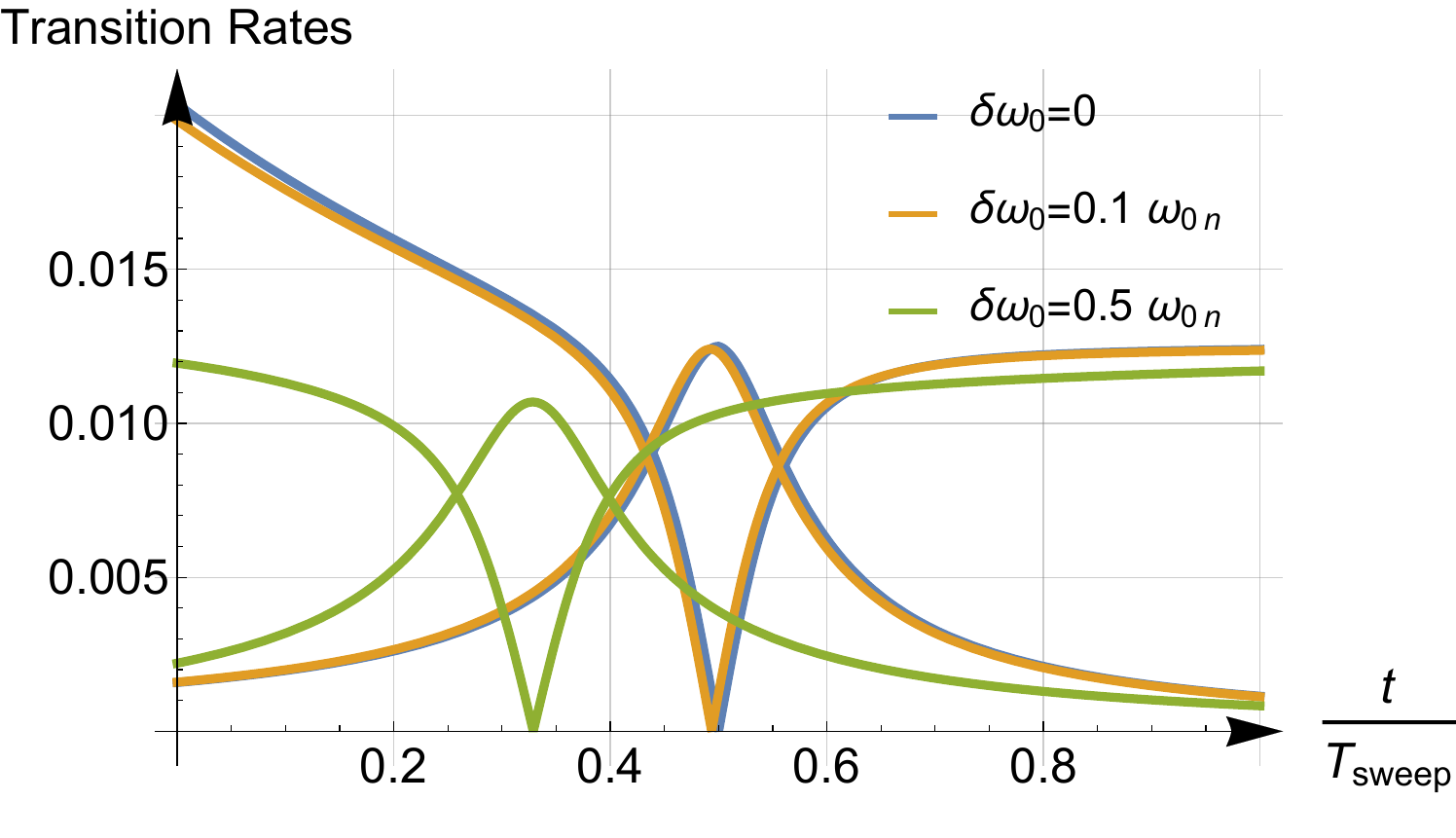}
  \caption[jsdjh]{ The transition rates between the instantaneous eigenstates of the $DQ$ subspace and that of the $ZQ$ subspace are plotted during the adiabatic sweep of the Rabi field for $A=0.1 \omega_{0n}$ and $C=0$. These leakage rates are considerably small even in the presence of a relative large local magnetic field such as $\delta \omega_{0}= 0.5\ \omega_{0n}$.}
  \label{Fig_rates}
  \end{figure}

In case of adiabatic-NOVEL, the above relations are a set of time dependent transition rates. We numerically evaluate them by replacing $\omega_{1e} \rightarrow \omega_{1e}(t)$ from Eq.\ref{Eq_transition_rates_2} and plot the result in Fig.\ref{Fig_rates}. We observe that right at the crossing point, where the electron energy matches the nuclear energy, the leakage between $|\widetilde{\Psi}_{\pm} \rangle \leftrightarrow |\widetilde{\Phi}_{\mp} \rangle $ is suppressed and$|\widetilde{\Psi}_{\pm} \rangle \leftrightarrow |\widetilde{\Phi}_{\pm} \rangle$ is maximized. We also observe that even for a large local field such as $\delta\omega_{0}= 0.5\ \omega_{0n}$, the leakage rate between the $DQ$ and the $ZQ$ is still very small but the cross polarization point is shifted to some time earlier as we expect.
  
With further increase of the local field (e.g., $C=\omega_{0n}$ in Figure \ref{Fig_VaryC}), at some point the reduction the polarization leakage is not negligible anymore and the scrambling of population between the $DQ$ and the $ZQ$ subspaces significantly affects the net enhancement. But, for the MRFM application, a local interaction as strong as $\omega_{0n} = 51$ MHz is very unlikely to happen. Therefore, for high field implementation of adiabatic-NOVEL, this DNP technique is highly robust and highly efficient. 

\section{Diffusion} 
\label{Sec_Diffusion}
In previous sections, we analyzed the polarization enhancement during the first round of the cross polarization phase. In this section, we include the spin diffusion phase and discuss the enhancement rate in subsequent iteration of the experiment. 

At the end of the spin locking period, the microwave field is off for a relatively long time so that the electron is reinitialized to the Boltzmann state, $T_{\text{off}}\gg T_{1e}$. During this time, the transferred polarization to the nuclear cloud diffuses to the other protons inside the sample by means of dipole-dipole interaction. Given the average distance between any two protons inside the sample, $a$, and given the spin decoherence time, $T_{2}$, the diffusion constant is estimated as $D\cong \frac{a^{2}}{50 T_{2}}$ \cite{Bloem49}. For water, $a= (\rho_{n})^{-1/3}\simeq 0.32 $ nm and the decoherence time can be upper bounded by the time constant of the flip-flop transition, which is estimated as $T_{dip} =\frac{ 1}{|H_{dip}|}\simeq 0.2 $ ms. Thus, $D\simeq1.8 \times 10^{-14}\  cm^{2}/s$.  Therefore, it takes approximately $T_{\text{diffus}} \simeq 100\  s$ for a local polarization at the surface of the nano-wire, that is induced during the cross polarization phase, to distribute over the bulk and reach the end of the sample at~$50$ nm away from surface. 

Consider an ideal situation where in the first round of the adiabatic-NOVEL, the electron is fully depolarized and that polarization is transferred to a cloud of $k$ nuclear spins near the surface.  For the second round, if there is no spin diffusion, the entropy difference between the fully polarized electron and the partially polarized nuclear cloud decreases. Therefore, the rate of polarization transfer to the cloud in subsequent rounds decrease monotonically. 

 In this case, the polarization of an average nuclear spin inside the cloud would reach the electron's polarization only in the asymptotic limit and the bulk is not polarized at all. In the other extreme limit, when the diffusion is almost instantaneous, the local polarization of the cloud distributes across the sample almost immediately. Under this unrealistic assumption, after a few iterations when the bulk obtains considerable polarization, the rate of information flow from the cloud to the bulk decreases monotonically. Thus, similar to the case of zero diffusion, an average nuclear spin inside the cloud would be fully polarized only in the asymptotic iteration of the experiment. In practice, the diffusion rate is neither instantaneous nor zero. Moreover, the number of times we repeat the experiment, $N^{*}$, is limited by nuclear relaxation time, because we require the polarized sample last long enough in order to do imaging afterwards. Therefore, the accumulated nuclear polarization during $N^{*}$ repetition is
\begin{eqnarray}
\langle P_{\text{cloud}}\rangle&=&\lim\limits_{N \rightarrow N^{*}} \ \sum\limits_{r=1}^{N}  \ \frac{1}{N}\ \bar{P}_{\text{cloud}}(r) < p_{e}\\ \nonumber
\end{eqnarray}  
where $\bar{P}_{\text{cloud}}(r)$ represents the average nuclear polarization of protons inside the cloud at the $r^{\text{th}}$ round and $p_{e}$ is electron thermal polarization. 

 \begin{figure}[h]
 \includegraphics[width=0.9\linewidth]{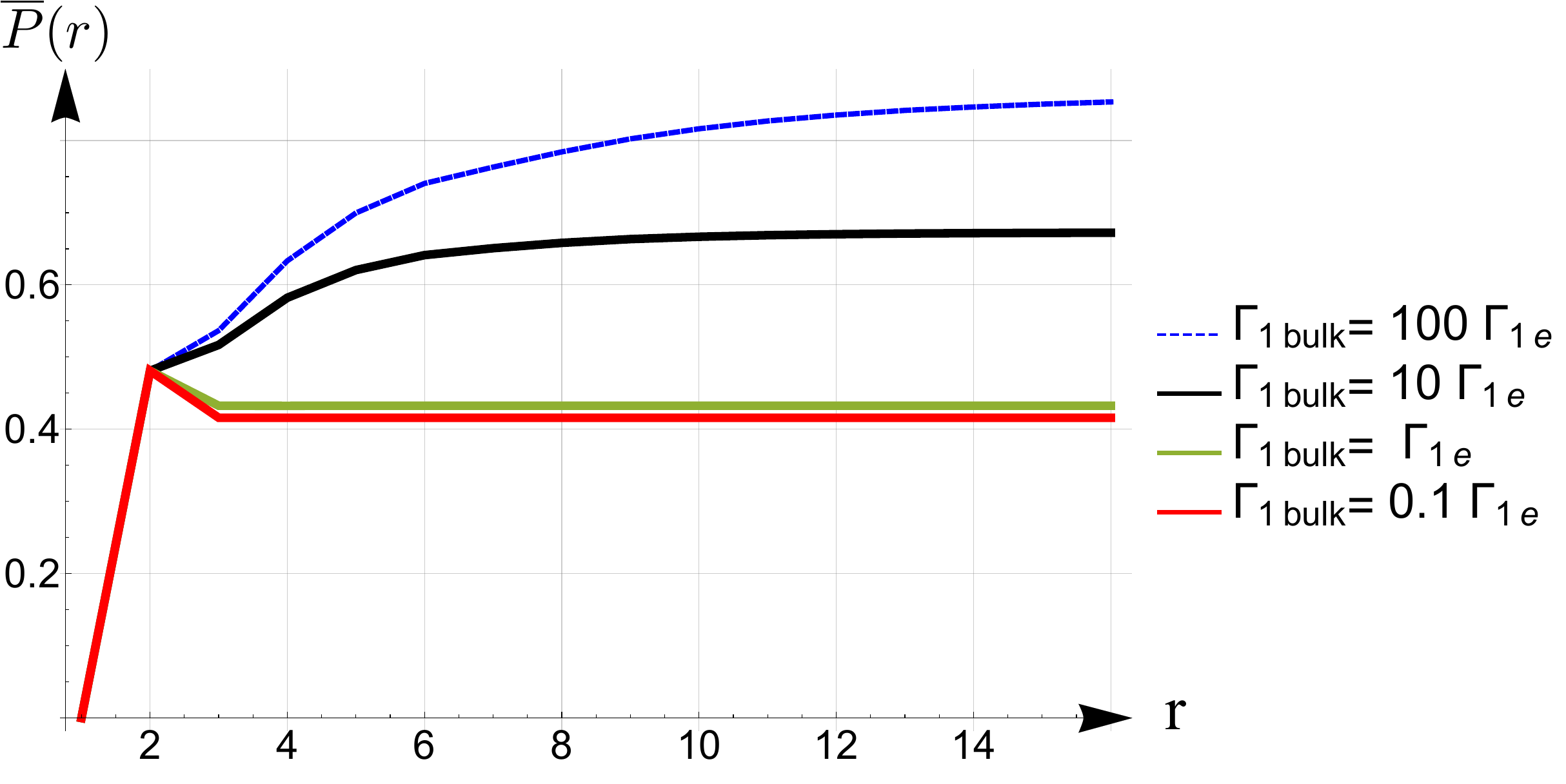}
 \caption{\label{Fig_PlotCloud} The average nuclear polarization of protons inside the cloud, as a function of experiment iteration, $r$. }
 \end{figure}

The Figure.\ref{Fig_PlotCloud} simulates the above discussion. We considered two relaxation processes in addition to the coherent dynamics: One that polarizes the electron at rate $(T_{1e})^{-1}=\Gamma_{1e}$ and another than depolarizes the nuclear spins in the cloud at rate $\Gamma_{1\text{bulk}}$. The former models the electron's relaxation process that pump out the entropy from the system. The latter models the flow of information from the cloud to the bulk which depends on both the diffusion rates and the nuclear relaxation time. The result shows that the enhancement rate in subsequent iteration of the experiment is limited by $\frac{\Gamma_{1e}}{\Gamma_{1\text{bulk}} }$.

\section{Conclusion}
\label{Sec_Conclusion}
In this work, we proposed an adiabatic version of nuclear orientation spin locking as a coherent DNP technique to hyperpolarize a nano-scale ensemble of nuclear spins. We analyzed the spin dynamics of an electron-nuclear under pulsed microwave irradiation. We compared the nuclear polarization enhancement for two cases of a constant Rabi field versus and a time-varying Rabi field. The result shows that with an adiabatic amplitude modulation of the microwave field, one can significantly improve the efficiency of NOVEL. Furthermore, we showed that the adiabatic -NOVEL is very robust against variation in local magnetic field. In particular, a deviation from the energy matching condition acts as a shift in the cross polarization time. For proof-of principles, we considered a linear amplitude modulation but one can extend this idea for both amplitude and frequency modulation with a non-linear time dependency.

We propose to use adiabatic-NOVEL as a sample preparation step for MRFM imaging to significantly enhance the image resolution. Adiabatic-NOVEL is fast, efficient and robust against field in-homogeneity, and therefore, is ideal for this particular application. Furthermore, the state-of-the art imaging device can operate at high static field, at low temperature and generates very strong microwave field with high precision of field control that are all ideal for implementing adiabatic-NOVEL. The presented theoretical analysis are in general but the numerical simulations were done for a particular MRFM set-up.

\begin{acknowledgments}
This  work  was  supported  by  the  US  Army  Research Office through Grant No.  W911NF-16-1-0199 . The authors would like to acknowledge the support from Canadian Excellence Research Chairs (CERC), Canada First  Research Excellence Fund (CFREF), the Natural  Sciences and Engineering  Research Council of Canada (NSERC), the Canadian Institute for Advanced Research and the Province of Ontario and Industry Canada.
\end{acknowledgments}

\appendix
\section{Adiabatic Rotation}
\label{Sec_AHP}
For the analysis of the cross polarization phase in section \ref{Sec_theory}, we assumed that the electron is initially prepared in $|\uparrow \rangle_{X}$ state which can be implemented by applying a $\frac{\pi}{2}$ rotation pulse on a thermally polarized electron. However, the current-carrying constriction generates a large microwave field inhomogeneity, and so, a hard pulse is not able to rotate all electron are rotated to the $X-Y$ plane. Therefore, we use an \textit{adiabatic half passage} pulse in order to rotate all the electrons across the sample to the $X$ direction \cite{H12, G01}. 

 We choose a non-linear adiabatic sweep where both the amplitude and phase of the microwave field are modulated. In particular, the hyperbolic functions are chosen because they are robust against both $B_{0}$ and $B_{1}$ inhomogeneities \cite{G01}. We consider

\begin{eqnarray}
\label{Eq_non_linear_sweep}
\omega_{1e}(s) &= & \omega_{\text{1max}} \ \sech[ \beta (s-1)], \\ \nonumber
\omega_{c}(s)  &=& \omega_{0e} + \alpha \ \tanh[ \beta ( s-1)] ,\\ \nonumber
\end{eqnarray}
where $\omega_{0e}$ is the center frequency of the electron spectrum, $s=t/T_{s}$ is a dimensionless parameter with $T_{s}$ being th total time of the adiabatic rotation and the parameter $\beta$ is chosen as $\sech[\beta] = 1- \tanh[\beta] = 0.01$ to avoid singularities. \\

  In order to satisfy the adiabaticity condition, the slope of changing the angle of the effective field in the rotating frame must be much smaller than the amplitude of the effective field in all instances of evolution. Therefore,

\begin{eqnarray}
\frac{d \theta (t) }{dt} &\ll & \sqrt{(\omega_{0e}- \omega_{c}(t))^2+ \omega_{1e}(t)^2} \\ \nonumber
&& \text{where}\\ \nonumber
 \theta(t) &=& \arctan[\frac{\omega_{1e}(t)}{(\omega_{0e}- \omega_{c}(t))}]\\ \nonumber
\end{eqnarray}
By replacing Eq.\ref{Eq_non_linear_sweep} into the above relation we obtain
\begin{eqnarray}
   \frac{1}{T_{s}} &\ll&\frac{1}{\beta} \ \frac{\omega_{\text{AHP}}^{3}}{\omega_{\text{1max}} \ \alpha} \ \cosh[\beta (s-1)]\ \nonumber
\end{eqnarray}
where $\omega_{\text{AHP}}(t)= \sqrt{(\omega_{0e}- \omega_{c}(t))^2+ \omega_{1e}(t)^2} \geq \omega_{\text{1max}}$. In order to optimize the efficiency of the adiabatic-NOVEL, it is important to us to choose $\omega_{\text{1max}}$ wisely so that during the electron rotation, there is almost no polarization exchange between the electron and the nuclear spin. Therefore, we want $\omega_{\text{AHP}}$ to be reasonably larger than the nuclear frequency, $\omega_{0n}$ at all times during the sweep. Thus, $\omega_{1max}$ should be reasonably larger than $\omega_{0n}$. The lower bound on the sweeping rate can be obtained by replacing the minimum value of the effective field and using $ \cosh[x]\geq1$. Thus, for $\omega_{\text{1max}} \geq 150\ \text{MHz}$ we have
\begin{equation}
T_{s}> \beta \frac{\alpha}{ \omega_{1max}^2} \approx 3 \ ns.
\end{equation}

\bibliographystyle{unsrt}
\renewcommand\bibname{References}
\bibliography{Bibliography_DNP_nano_MRI_2.bib}

\begin{thebibliography}{10}

\bibitem{A62}
{A. Abragam}.
\newblock {\em The principles of nuclear magnetisim}.
\newblock Clarendon Press, Oxford, 1962.

\bibitem{S96}
{C. P. Slichter}.
\newblock {\em Principles of magnetic resonance}.
\newblock Springer, 1962.

\bibitem{S65}
T.~J. Schmugge and C.~D. Jeffries.
\newblock High dynamic polarization of protons.
\newblock {\em Phys. Rev.}, 138:A1785--A1801, 1965.

\bibitem{Over53}
A.W. Overhauser.
\newblock Polarization of nuclei in metals.
\newblock {\em Phys. Rev.}, 92:411--415, 1953.

\bibitem{CS53}
{T.R. Carver, C.P. Slichter}.
\newblock {Polarization of nuclear spins in metals}.
\newblock {\em Phys. Rev.}, 92:212--213, 1953.

\bibitem{Gol06}
{Golman, Klaes and Zandt, Ren{\'e} in{\textquoteright}t and Lerche, Mathilde
  and Pehrson, Rikard and Ardenkjaer-Larsen, Jan Henrik}.
\newblock {Metabolic Imaging by Hyperpolarized 13C Magnetic Resonance Imaging
  for In vivo Tumor Diagnosis}.
\newblock {\em Cancer Research}, 66(22):10855--10860, 2006.

\bibitem{Gol062}
{Golman,Klaes and Petersson, J. Stefan }.
\newblock {Metabolic Imaging and Other Applications of Hyperpolarized 13C1}.
\newblock {\em Academic Radiology}, 13(8):932 -- 942, 2006.

\bibitem{Sche17}
J.~Scheuer and et~al.
\newblock Robust techniques for polarization and detection of nuclear spin
  ensembles.
\newblock 2017.

\bibitem{Lon13}
P.~London and et~al.
\newblock Detecting and polarizing nuclear spins with double resonance on a
  single electron spin.
\newblock {\em Physical Revie Letter}, 111, 2013.

\bibitem{A09}
J.~W. Aptekar and et~al.
\newblock Silicon nanoparticles as hyperpolarized magnetic resonance imaging
  agents.
\newblock {\em ACS Nano}, 3(12):4003--4008, 2009.

\bibitem{Nelson08}
J.~Kurhanewicz A. Chen R.~Bok S.~J.~Nelson, D.~Vigneron and R.~Hurd.
\newblock Dnp-hyperpolarized ${13}^c$ magnetic resonance metabolic imaging for
  cancer applications.
\newblock {\em Appl Magn Reson}, 34(3-4):533–544, 2008.

\bibitem{M08}
Thorsten Maly, Galia~T. Debelouchina, Vikram~S. Bajaj, Kan-Nian Hu, Chan-Gyu
  Joo, Melody L.~Mak? Jurkauskas, Jagadishwar~R. Sirigiri, Patrick C.~A.
  van~der Wel, Judith Herzfeld, Richard~J. Temkin, and Robert~G. Griffin.
\newblock Dynamic nuclear polarization at high magnetic fields.
\newblock {\em J. Chem. Phys.}, 128:052211, 2008.

\bibitem{T08}
{Taylor, J. M. and Cappellaro, P. and Childress, L. and Jiang, L. and Budker,
  D. and Hemmer, P. R. and Yacoby, A. and Walsworth, R. and Lukin, M. D.}
\newblock {High-sensitivity diamond magnetometer with nanoscale resolution}.
\newblock {\em Nature Physics}, 4, 2008.

\bibitem{MM10}
{Maurer, P. C. and et al.}
\newblock {Far-field optical imaging and manipulation of individual spins with
  nanoscale resolution}.
\newblock {\em Nature Physics}, 6, 2010.

\bibitem{Ru04}
{Rugar, D. and Budakian, R. and Mamin, H. J. and Chui, B. W.}
\newblock {Single spin detection by magnetic resonance force microscopy}.
\newblock {\em Nature}, 430(2), 2004.

\bibitem{De09}
C.~L. Degen, M.~Poggio, H.~J. Mamin, C.~T. Rettner, and D.~Rugar.
\newblock Nanoscale magnetic resonance imaging.
\newblock {\em Proceedings of the National Academy of Sciences},
  106(5):1313--1317, 2009.

\bibitem{N13}
John~M. Nichol, Tyler~R. Naibert, Eric~R. Hemesath, Lincoln~J. Lauhon, and
  Raffi Budakian.
\newblock Nanoscale fourier-transform magnetic resonance imaging.
\newblock {\em Phys. Rev. X}, 3:031016, 2013.

\bibitem{B99}
{ Simon J. BENDING }.
\newblock {Local magnetic probes of superconductors}.
\newblock {\em Advances in Physics}, 48:449--535, 1999.

\bibitem{KL04}
{ Kleiner, R. and Koelle, D. and Ludwig, F. and Clarke, J. }.
\newblock {Superconducting quantum interference devices: State of the art and
  applications.}
\newblock {\em Proc. IEEE}, 92:1534–1548, 2004.

\bibitem{Sa11}
{Sawicki, M. and Stefanowicz, W. and Ney, A.}
\newblock {Sensitive SQUID magnetometry for studying nanomagnetism}.
\newblock {\em Semiconductor Science and Technology}, 26(6):064006, 2011.

\bibitem{HS68}
{K.H. Hausser, D. Stehlik}.
\newblock {Dynamic nuclear polarization in liquids}.
\newblock {\em Adv. Magn. Reson.}, 3:79, 1968.

\bibitem{P73}
A.~Pines, M.~G. Gibby, and J.~S. Waugh.
\newblock Proton enhanced nmr of dilute spins in solids.
\newblock {\em J. Chem. Phys.}, 59:569--590, 1973.

\bibitem{M79}
L.~Müller and R.R. Ernst.
\newblock Coherence transfer in the rotating frame.
\newblock {\em Molecular Physics}, 38(3):963--992, 1979.

\bibitem{W06}
{V. Weis and R.G. Griffin}.
\newblock {Electron-nuclear cross polarization }.
\newblock {\em {Solid State Nuclear Magnetic Resonance}}, 29:66 -- 78, 2006.

\bibitem{Sara13}
S.~Sheldon.
\newblock {\em Optimal Control in an open quantum system: selecting DNP
  Pathways in an electron-nuclear system}.
\newblock Massachuestts Institite of Technology, Ph. D. thesis, 2013.

\bibitem{YFS10}
S.~Vega. Y.~Hovav, A.~Feintuch.
\newblock Theoretical aspects of dynamic nuclear polarization in the solid
  state – the solid effect.
\newblock {\em Journal of Magnetic Resonance}, 207:176–189, 2010.

\bibitem{H62}
S.~R. Hartmann and E.~L. Hahn.
\newblock Nuclear double resonance in the rotating frame.
\newblock {\em Phys. Rev.}, 128:2042--2053, 1962.

\bibitem{HDSW88}
A~Henstra, P~Dirksen, J~Schmidt, and W.Th Wenckebach.
\newblock Nuclear spin orientation via electron spin locking (novel).
\newblock {\em J. Mag. Res.}, 77:389--393, 1988.

\bibitem{B87}
H.~Brunner, R.~H. Fritsch, and K.~H. Hausser.
\newblock Cross polarization in electron nuclear double resonance by satisfying
  the hartmann-hahn condition.
\newblock {\em Z. Naturforsch}, 42a:1456--1457, 1987.

\bibitem{H08}
A.~Henstra and W.Th. Wenckebach.
\newblock The theory of nuclear orientation via electron spin locking (novel).
\newblock {\em Molecular Physics}, 106(7):859--871, 2008.

\bibitem{SGBRZ95}
K.~J. Bruland D. Rugar O. Zuger S.~Hoen J.~A.~Sidles, J. L.~Garbini and C.~S.
  Yannoni.
\newblock Magnetic resonance force microscopy.
\newblock {\em Rev. Mod. Phys.}, 67:249, 1995.

\bibitem{P57}
D.~Pines, J.~Bardeen, and C.~P. Slichter.
\newblock Nuclear polarization and impurity-state spin relaxation processes in
  silicon.
\newblock {\em Phys. Rev.}, 106:489--498, 1957.

\bibitem{Pro62}
B.~N. Provotqrov.
\newblock Theory of double magnetic resonance in solids.
\newblock {\em Physical Review}, 128, 1962.

\bibitem{Can15}
T.~M.~Swager T.~V.~Can, J. J.~Walish and R.~G. Griffin.
\newblock Time domain dnp with the novel sequence.
\newblock {\em The Journal of Chemical Physics}, 143, 2015.

\bibitem{Can17}
T.~V. Can and et~al.
\newblock Ramped-amplitude novel.
\newblock {\em The journal of chemical physics}, 146:154204, 2017.

\bibitem{N12}
John~M. Nichol, Eric~R. Hemesath, Lincoln~J. Lauhon, and Raffi Budakian.
\newblock Nanomechanical detection of nuclear magnetic resonance using a
  silicon nanowire oscillator.
\newblock {\em Phys. Rev. B}, 85:054414, Feb 2012.

\bibitem{ME79}
L.~Muller and R.~R. Ernst.
\newblock Coherence transfer in the rotating frame application to heteronuclear
  cross-correlation spectroscopy.
\newblock {\em Molecular Physcs}, 38, 1979.

\bibitem{D08}
D.~G.~Cory Dementyev, E.~Anatoly and R.~Chandrasekhar.
\newblock Rapid diffusion of dipolar order enhances dynamic nuclear
  polarization.
\newblock {\em Phys. Rev. B}, 77:024413, 2008.

\bibitem{S85}
{D. Suter and R.R. Ernst}.
\newblock {Spin diffusion in resolved solid-state NMR spectra}.
\newblock {\em Phys. Rev. B}, 32:5608--5627, 1985.

\bibitem{Z98}
Wurong Zhang and D.~G. Cory.
\newblock First direct measurement of the spin diffusion rate in a homogenous
  solid.
\newblock {\em Phys. Rev. Lett.}, 80:1324--1327, 1998.

\bibitem{Ho11}
Yonatan Hovav, Akiva Feintuch, and Shimon Vega.
\newblock Dynamic nuclear polarization assisted spin diffusion for the solid
  effect case.
\newblock {\em J. Chem. Phys.}, 134:074509, 2011.

\bibitem{Ram08}
C.~Ramanathan.
\newblock Dynamic nuclear polarization and spin diffusion in nonconducting
  solids.
\newblock {\em Appl Magn Reson}, 34:409--421, 2008.

\bibitem{GA01}
{Michael Garwood and Lance DelaBarre}.
\newblock {The Return of the Frequency Sweep: Designing Adiabatic Pulses for
  Contemporary NMR}.
\newblock {\em Journal of Magnetic Resonance}, 153(2):155 -- 177, 2001.

\bibitem{B89}
R.~Biswas, C.~Z. Wang, C.~T. Chan, K.~M. Ho, and C.~M. Soukoulis.
\newblock Electronic structure of dangling and floating bonds in amorphous
  silicon.
\newblock {\em Phys. Rev. Lett.}, 63:1491--1494, 1989.

\bibitem{S07}
Rogerio de~Sousa.
\newblock Dangling-bond spin relaxation and magnetic $1?f$ noise from the
  amorphous-semiconductor/oxide interface: Theory.
\newblock {\em Phys. Rev. B}, 76:245306, 2007.

\bibitem{WY07}
{Y. Wu and X. Yang}.
\newblock {Strong-Coupling Theory of Periodically Driven Two-Level Systems}.
\newblock {\em Phys. Rev. Lett.}, 98:013601, 2007.

\bibitem{Zener32}
C.~Zener.
\newblock Non-adiabatic crossing of energy levels.
\newblock {\em Proc. R. Soc. A}, 137:696--702, 1932.

\bibitem{Bloem49}
N.~Bloembergen.
\newblock On the interaction of nuclear spins in a crystalline lattice.
\newblock {\em Physica}, 15:386--426, 1949.

\bibitem{H12}
Kristopher~J. Harris, Adonis Lupulescu, Bryan~E.G. Lucier, Lucio Frydman, and
  Robert~W. Schurko.
\newblock Broadband adiabatic inversion pulses for cross polarization in
  wideline solid-state nmr spectroscopy.
\newblock {\em J. Mag. Res.}, 224:38--47, 2012.

\bibitem{G01}
{M. Garwood and L. DelaBarre}.
\newblock {The Return of the Frequency Sweep: Designing Adiabatic Pulses for
  Contemporary NMR}.
\newblock {\em J Mag. Res.}, 153:155 -- 177, 2001.

\end{thebibliography}


\end{document}